\documentclass[
 reprint,
superscriptaddress,
 amsmath,amssymb,
 aps,
prd,
floatfix,
]{revtex4-2}
\usepackage[dvipsnames]{xcolor}
\usepackage{graphicx}%
\usepackage{dcolumn}%
\usepackage{bm}%
\usepackage{siunitx}
\usepackage{xcolor}
\usepackage{xspace}
\usepackage{booktabs}
\usepackage{multirow}
\usepackage{physics}
\usepackage{makecell}
\usepackage[breaklinks=true]{hyperref}
\usepackage[normalem]{ulem}
\usepackage{cleveref}
\usepackage{afterpage}

\makeatletter
\let\MYcaption\@makecaption
\makeatother

\usepackage{subcaption}

\makeatletter
\let\@makecaption\MYcaption
\makeatother

\newcommand{\decayp}{\SI{459}{MeV/\textit{c}}\xspace}
\newcommand{\gibuu}{\textsc{GiBUU}\xspace}
\newcommand{\geant}{\textsc{Geant4}\xspace}
\newcommand{\nuwro}{\textsc{NuWro}\xspace}
\newcommand{\genie}{\textsc{GENIE}\xspace}
\newcommand{\neut}{\textsc{NEUT}\xspace}

\newcommand{\nucleon}{\textrm{N}}

\newcommand{\wcsim}{\textsc{WCSim}\xspace}

\newcommand{\figref}[1]{Fig.~\ref{#1}}
\newcommand{\gibver}{25p3\xspace}
\newcommand{\cda}{CdA\xspace}
\newcommand{\positron}{\textrm{e}^+}
\newcommand{\proton}{\textrm{p}}
\newcommand{\pdk}{\proton \to \positron \pi^0}
\newcommand{\patmo}{P_\textrm{atmo}}
\newcommand{\pbound}{P_\textrm{bound}}
\newcommand{\pfree}{P_\textrm{free}}
\newcommand{\cfree}{\textrm{C}_\textrm{free}}
\newcommand{\catmo}{\textrm{C}_\textrm{atmo}}
\newcommand{\cbound}{\textrm{C}_\textrm{bound}}
\newcommand{\ggonly}{{\gamma\gamma}}
\newcommand{\egg}{{\positron\ggonly}}
\newcommand{\pegg}{p_\egg}
\newcommand{\megg}{M_\egg}
\newcommand{\pgg}{p_\ggonly}
\newcommand{\thetaegg}{\theta_{\positron\textrm{-}\gamma\gamma}}
\newcommand{\catag}{{\rm C}_{i}}
\newcommand{\Bayesp}{\mathcal{P}}
\newcommand{\priorp}{\Bayesp}
\newcommand{\poissp}{\mathbb{P}}
\newcommand{\los}{1}%
\newcommand{\ups}{2}%

\newcommand{\cll}{\textrm{90}}
\newcommand{\glim}{\overline{\Gamma}_\cll}%
\newcommand{\gexp}{\widehat{\Gamma}_\cll}
\newcommand{\texp}{\widehat{\tau}_\cll}

\newcommand{\figwid}{\linewidth}
\definecolor{darkgreen}{rgb}{0.0, 0.5, 0.0}
\definecolor{darkblue}{rgb}{0.0, 0.0, 0.7}

\newcommand{\warwick}{University of Warwick, Department of Physics, Coventry, CV4 7AL United Kingdom}
\newcommand{\ucas}{School of Physical Sciences, University of Chinese Academy of Sciences, Beijing 100049, China}
\newcommand{\sysu}{Sino-French Institute of Nuclear Engineering and Technology, Sun Yat-sen University, Zhuhai, China}
\newcommand{\giessen}{Institut f\"ur Theoretische Physik, Universit\"at Giessen, 35392 Giessen, Germany}

\crefname{section}{Sec.}{Secs.}
\crefname{subsection}{Sec.}{Secs.}
\crefname{figure}{Fig.}{Figs.}
\crefname{table}{Table}{Tables}
\crefname{equation}{Eq.}{Eqs.}
\creflabelformat{equation}{#2#1#3}

\begin{document}
\title{Understanding the impact of nuclear effects on proton decay searches\\ with the GiBUU model}%

\author{Qiyu Yan}
\affiliation{\ucas}%
\affiliation{\warwick}%

\author{Akira Takenaka}
\email{Corresponding author: takenakaa@mail.sysu.edu.cn}
\affiliation{\sysu}

\author{Kai Gallmeister}
\altaffiliation[Present address: ]{Q-Leap Networks, 75378 Bad Liebenzell, Germany}
\affiliation{\giessen}

\author{Xianguo Lu}
\affiliation{\warwick}

\author{Ulrich Mosel}
\affiliation{\giessen}

\author{Yangheng Zheng}
\affiliation{\ucas}%

\date{\today}%

\begin{abstract}

    Proton decay searches in the next generation of water Cherenkov detectors, such as Hyper-Kamiokande, are expected to probe the $10^{35}$-year lifetime regime where atmospheric neutrino backgrounds and systematic uncertainties begin to play an increasingly important role.
    In this study, we employ the \gibuu framework and reevaluate the proton decay search sensitivity for the $\pdk$ channel by incorporating a typical event reconstruction performance in water Cherenkov detectors.
    Using sophisticated models implemented in \gibuu---most notably the mean-field potential and Boltzmann transport---which have been benchmarked against accelerator neutrino scattering data, in particular pion production, we find that the resulting proton decay signal detection efficiency and atmospheric neutrino background rate are comparable to those previously evaluated for the current and near future water Cherenkov experiments using \textit{ad hoc} nuclear models.
    In addition to pion final-state interactions, we evaluate the impact of differences in the Fermi momentum distribution of nucleons in the nucleus, as a source of systematic uncertainty, on the signal detection efficiency and the expected background event rate.
    We find that the uncertainty associated with pion final-state interactions is moderate, whereas the choice of Fermi momentum distribution can significantly affect the estimated atmospheric neutrino background rate and constitutes the dominant contribution.
    Our study provides an independent and complementary characterisation of nuclear effects on proton decay searches and helps to refine sensitivity estimates in the regime where systematic uncertainties become more relevant.

\end{abstract}

\maketitle

\section{Introduction}\label{sec:intro}
Grand Unified Theories (GUTs) are among the most compelling extensions of the Standard Model of particle physics, providing a framework to unify the strong, weak, and electromagnetic forces into a single interaction at super-high energy scales~\cite{Georgi:1974sy, Langacker:1980js, Georgi:1974yf, Fritzsch:1974nn, Gursey:1975ki}.
By incorporating leptons and quarks into unified multiplets and introducing heavy gauge bosons to mediate interactions between them, GUTs inherently predict the instability of the proton.

As a unique prediction from many GUTs, the detection of proton decay would offer profound insights into the fundamental nature of matter and the forces that govern the universe.
Among the various predicted decay modes, the channels involving a positron accompanied by a neutral pion ($\pdk$) are particularly noteworthy due to their relatively high branching ratios in many of the proposed GUT models and their clear experimental signatures.

This decay mode has been and will be targeted by many experiments, including IMB~\cite{McGrew:1999nd}, Kamiokande~\cite{Kamiokande-II:1989avz}, Super-Kamiokande~\cite{Super-Kamiokande:1998mae, Super-Kamiokande:2009yit, Super-Kamiokande:2016exg, Super-Kamiokande:2020wjk}, and the upcoming Hyper-Kamiokande experiment~\cite{Hyper-Kamiokande:2018ofw}.
But current experimental results do not yet provide any positive evidence for proton decay, leading to stringent lower limits on the proton's lifetime.
Currently, the most stringent limits come from the Super-Kamiokande experiment, which has set lower bounds on the proton lifetime of $\tau / B(\pdk) > 2.4 \times 10^{34}$~years at a 90\% confidence level~\cite{Super-Kamiokande:2020wjk}, where $B(\pdk)$ is the decay branching ratio for the $\pdk$ mode.

Given the proton lifetime predicted by GUTs to exceed $10^{30}$~years, as a consequence of the super-heavy mass scale of the mediator bosons ($10^{14}$ to $10^{15}$~GeV scale), large-scale water Cherenkov detectors such as IMB, Kamiokande, and Super-Kamiokande were constructed in order to monitor a large number of protons in their water volume for the detection of proton decay signals.
These detectors are equipped with highly sensitive optical sensors, namely photomultiplier tubes, mounted on the detector walls to detect Cherenkov light emitted by particles produced after the decay.
Based on the number of detected photons and their detection times at individual photomultiplier tubes, the number of observed particles in the form of Cherenkov rings, their  identity and kinematics, including the directions, are reconstructed~\cite{Shiozawa:1999sd}.
In the most recent proton decay searches conducted by the Super-Kamiokande experiment~\cite{Super-Kamiokande:2020wjk}, the signal detection efficiency is estimated to be approximately 38\%.
The background event rate originating from atmospheric neutrinos is estimated to be around 1.8 events/(Mton$\cdot$years) without neutron tagging and 1.0 events/(Mton$\cdot$years) when background reduction using neutron detection is taken into account.
The loss in detection efficiency is primarily attributed to nuclear effects associated with proton decay in oxygen nuclei.
The Hyper-Kamiokande experiment~\cite{Hyper-Kamiokande:2018ofw}, whose detector with a fiducial mass of approximately 200~kton is currently under construction, is expected to achieve nearly comparable proton decay search performance in terms of the signal detection efficiency and background rates.
With 10 years of operation, the expected search sensitivity at the 90\% confidence level is projected to reach a proton lifetime of $10^{35}$~years. \par
Other large neutrino detectors such as JUNO~\cite{JUNO:2022qgr} and DUNE~\cite{DUNE:2020fgq} also possess high sensitivity to proton decay searches, particularly for the ${\proton}\rightarrow \nu {\rm K}^{+}$ mode.
However, the $\pdk$ mode has been searched for many years using water Cherenkov detectors, and its event reconstruction performance is therefore better understood.
In addition, Hyper-Kamiokande is expected to achieve the highest sensitivity for the $\pdk$ mode.
For these reasons, the present study focuses on discussing the proton decay search sensitivity of the Hyper-Kamiokande detector in the $\pdk$ channel.

\begin{figure}[htbp]
    \includegraphics[width=0.7\linewidth]{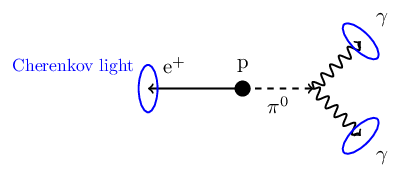}
    \caption{Schematic of the free proton decay, $\pdk$, in a water Cherenkov detector.
        The positron,  as well as the two photons from the $\pi^{0}$ decay, generate electromagnetic cascades in water and emit radially distributed Cherenkov light.}
    \label{fig:pdk_schematic}
\end{figure}

In water Cherenkov detectors, the decay of a hydrogen nucleus (free proton) in water molecules produces a final state consisting of a positron and a neutral pion emitted back-to-back, each carrying a monoenergetic momentum of  \decayp.
The neutral pion subsequently decays almost instantaneously ($\tau\sim10^{-16}$~sec) into two photons, which then produce electromagnetic showers detectable via Cherenkov radiation, leaving a 3-ring signature in the detector as demonstrated in Fig.~\ref{fig:pdk_schematic}.

However,  $80\%$ of the protons in water are bound in the oxygen nuclei.
The nuclear environment introduces several complexities that can significantly alter the expected signatures of proton decay.
Those nuclear effects include the Fermi motion (FM), which imparts an initial momentum to the decaying proton, leading to a smearing of the final state momenta;
and short-range correlations (SRCs), where the decaying proton may be correlated with another nucleon, affecting the kinematics of the decay products;
binding energy, which effectively reduces the available energy for the decay products;
and final-state interactions (FSIs), where the decay products may interact with other nucleons in the nucleus before exiting, potentially altering their 4-momenta or even changing the topology of the final state by producing additional particles, exchanging the electric charge, or absorbing some of the decay products.

These nuclear effects complicate the reconstruction of proton decay events and introduce a significant loss in detection efficiency, as well as increased systematic uncertainties in the analysis.
The understanding and modeling of those nuclear effects are crucial for accurately interpreting the results of proton decay searches, setting reliable limits on the proton lifetime, and evaluating the search sensitivity.

In this work, we reassess the impact of nuclear effects on the sensitivity of next-generation proton decay searches in the $\pdk$ channel. Using the \gibuu
model, we provide a consistent treatment of nuclear effects for both the proton decay signal and the atmospheric neutrino background.
This approach differs from many previous studies, which typically employ intranuclear cascade (INC)-based event generators such as \genie~\cite{Andreopoulos:2009rq},  \nuwro~\cite{Golan:2012wx}, and \neut~\cite{Hayato:2021heg}. These generators implement nuclear effects using factorized models of the initial state and cascade-based treatments of final-state interactions.
Our work thus offers a fresh perspective and a more unified modeling of the systematic uncertainties associated with nuclear effects in proton decay searches.

This paper is organized as follows.
Section~\ref{sec:nuclear_effects} discusses nuclear effects that influence proton decay signals and atmospheric neutrino backgrounds. Section~\ref{sec:gibuu} introduces the \gibuu model used in this study.
Section~\ref{sec:setup-and-analysis} presents the proton decay search sensitivity estimated using \gibuu and a water Cherenkov detector response model, followed by a summary of this paper in Sec.~\ref{sec:conclusion}.

\section{Nuclear effects in proton decay}\label{sec:nuclear_effects}

The proton decay signal is %
modified by various nuclear effects, including FM, binding energy, SRCs, and FSIs,  when the decay occurs within a bound nucleus, such as oxygen in water Cherenkov detectors.
The kinematics of the decay products can be significantly altered, making the signal identification and reconstruction more challenging compared to the idealized case of a free proton decay, where the decay products are back-to-back $\positron$ and $\pi^0$, each with a momentum of \decayp.

\subsection{Initial-state effects}\label{sec:initial_state}
The initial-state proton inside the nucleus is not at rest but possesses a momentum distribution due to FM. This intrinsic momentum distribution smears the final-state momenta of the decay products. %
The binding energy of the proton within the nucleus also plays a crucial role, effectively reducing the available energy and leading to lower momenta for the decay products compared to those in the free proton case, further complicating the reconstruction of proton decay events.

Theoretical modeling of these initial-state nuclear effects is essential for accurately predicting the kinematic distributions of the proton decay products. Various nuclear models, such as the local Fermi gas (LFG) model~\cite{Bethe:1968zza} or spectral function approaches~\cite{Benhar:1994hw}, can be employed to describe the momentum and energy distributions of nucleons within the nucleus.

In LFG, nucleons occupy states below the Fermi surface. However, both experimental and theoretical studies indicate that nucleon–nucleon correlations can significantly broaden the single-nucleon momentum distribution; this is the SRC effect.
Modeling SRC is challenging because it involves complex nuclear dynamics. When the decaying proton belongs to a correlated pair, there are substantial deviations from the expected kinematics of uncorrelated proton decay: the initial-state momentum of the proton can be significantly higher than the Fermi surface, and the decay kinematics are affected by the partner nucleon, which absorbs part of the momentum during the process.

In Ref.~\cite{Yamazaki:1999gz}, which is also referred to in the latest proton decay search in Super-Kamiokande~\cite{Super-Kamiokande:2020wjk},
the fraction of proton decays in oxygen nuclei that involve the SRC effect---refereed to as ``correlated decay'' there---was estimated to be approximately 10\%.
This nucleon–nucleon correlation
is accounted for by treating the decay as a three‑body process subject to Pauli blocking (requiring the final recoil nucleon momentum to exceed the Fermi momentum).
The decay occurs on a pair of correlated nucleons,
i.e. ${\proton\nucleon} \rightarrow \positron\pi^0{\nucleon}$, where $\nucleon$ denotes the partner nucleon.
The momentum of the decaying proton is sampled according to the customized Fermi momentum model established based on the electron scattering experiment~\cite{Nakamura:1976mb}.
The momentum transfer to the recoiling partner nucleon is based on the nucleon-nucleon correlation function obtained from the Reid soft-core potential~\cite{Reid:1968sq}.
In this treatment, the invariant mass of the initial correlated system is fixed to the total mass of the two nucleons,
without the binding energy effects~\footnote{In the  proton decay event generation in Super-Kamiokande, upon sampling with a given fraction, nucleon–nucleon correlation effects are incorporated through the introduction of an effective particle with a mass equal to twice the proton mass.
    This effective particle is allowed to decay into a proton, a positron, and a neutral pion thereby modeling the kinematic features associated with correlated two-nucleon configurations~\cite{Takenaka:2020cmb}.}.
The calculation indicates that the SRC‑induced proton decay produces a broad pseudo‑peak at a lower invariant mass~\cite{Takenaka:2020cmb}.

In this work, as an alternative, the
high-momentum tail well beyond the Fermi surface
has been parameterized by Cioffi degli Atti and Simula \cite{CiofidegliAtti:1995qe}. This momentum distribution is, as an option, implemented in \gibuu (see Sec.~\ref{sec:gibuu}) while the energy-distribution is taken to be that of a free Fermi-gas. In the following discussions, the use of this correlated momentum distribution is denoted by the abbreviation, \cda.

\subsection{Final-state interactions}\label{sec:final_state}
After the proton decay, $\pdk$, occurs,
the $\pi^0$ can undergo various FSIs with the surrounding nucleons before exiting the nucleus. These interactions include absorption, charge exchange, and scattering processes that can alter the momentum and final state topology of the decay products.

The rescattering process of the $\pi^0$ in the nucleus involves complicated mechanisms and possible medium modifications to the intermediate resonances, such as the $\Delta$ resonance that contributes dominantly to pion-nucleon interactions in the relevant energy range. The $\Delta$ resonance properties, including its mass and width, can be modified in the nuclear medium due to interactions with surrounding nucleons.

The effects of the Coulomb force on the $\positron$ outgoing from the nucleus have been studied in the context of neutrino interactions~\cite{Plestid:2023mta, Tomalak:2021hec, Tomalak:2022xup, Tomalak:2024lme} and are of theoretical interest.
However, since the Coulomb potential is of the order of a few~MeV and is sufficiently small compared to the energies of particles produced in proton decay, these effects are not taken into account in the present study.

\subsection{Background contamination} \label{sec:background}

Atmospheric neutrino interactions constitute the dominant and irreducible background in proton decay searches because they can produce final-state kinematics and topologies that closely mimic genuine proton decay signals.
Both proton decay and atmospheric-neutrino interactions are subject to common initial-and final-state nuclear effects which modify the observable kinematics and topology~\cite{Mosel:2023zek,bogart2024inmedium,Yan:2025aau,Yan:2024kkg,GENIE:2024ufm}. In the case of atmospheric neutrinos, however, an additional and intrinsically distinct ingredient enters: the primary neutrino–pion production.
Recent studies exploring alternative treatments of resonant and non-resonant pion production within \nuwro~\cite{Gonzalez-Jimenez:2016qqq,Nikolakopoulos:2022tut,Niewczas:2020fev,Golan:2012wx,Yan:2024kkg}  indicate potential non-negligible variations in predicted atmospheric-neutrino pion production. A similar study for an alternative variation in the neutrino-pion production was done using \gibuu~\cite{Yan:2025aau} (details in Sec.~\ref{sec:gibuu}).

In contrast to event-generator frameworks such as \nuwro and \genie, which typically treat the primary interaction and final-state interactions within a modular structure, \gibuu employs an approach in which both stages are handled consistently within the same dynamical framework,
and we will therefore focus on the theoretical systematic uncertainties evaluated within \gibuu in this work.

\subsection{Ternary classification}\label{role_of_nuclear_effects}

\begin{figure}[!htbp]
    \includegraphics[width=\figwid]{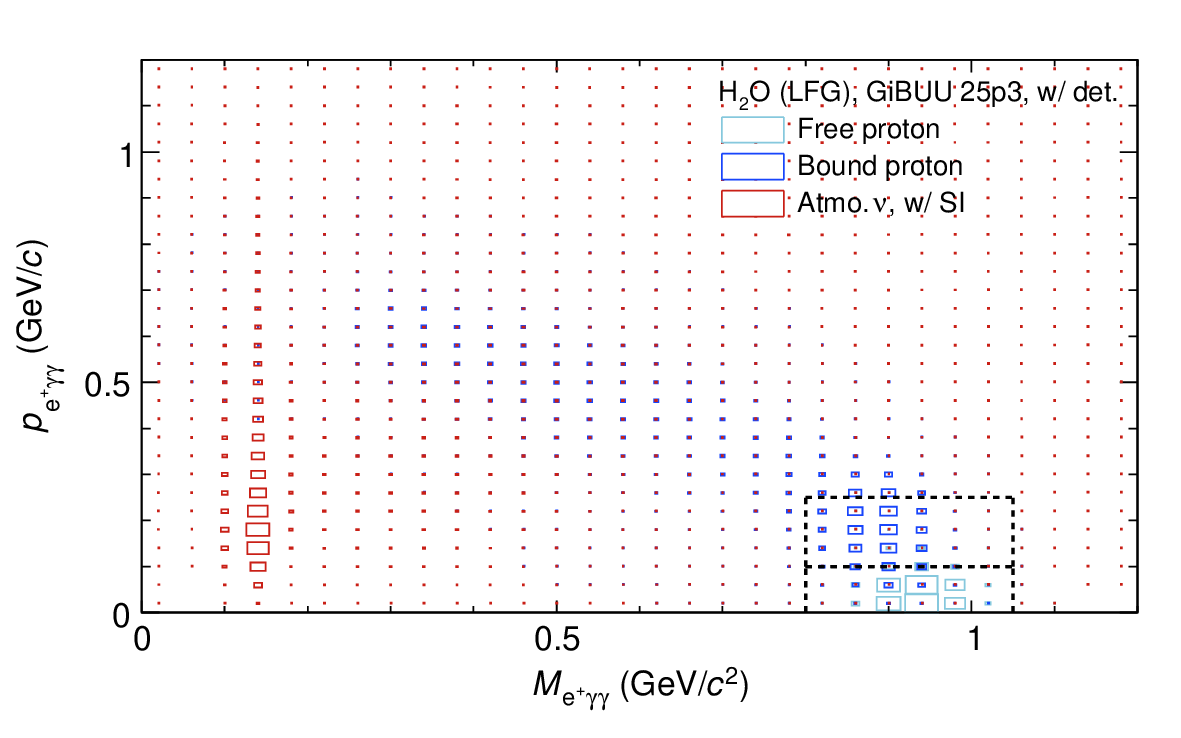}
    \caption{Momentum ($\pegg$) vs. invariant mass ($\megg$) of the proton decay candidates. True events from different sources---free proton decay, bound proton decay in oxygen, and atmospheric neutrino interactions---are considered.
        All atmospheric neutrino events are simulated for interactions on a water target. The nuclear initial state is modeled  with local Fermi gas (LFG).
        The box size corresponds to the normalized event population, shown on a linear scale.
        The treatment of nuclear effects in \gibuu is described in Sec.~\ref{sec:gibuu}.
        Detector effects (det.) are modeled based on the Super-Kamiokande reconstruction performance as described in Sec.~\ref{sec:detector_modeling} with secondary interactions (SIs) to be discussed in Sec.~\ref{sec:eventgen_atm}.
        The signal region defined in the latest Super-Kamiokande analysis, which is also adopted in the present analysis  (see \textbf{C5} and \textbf{C6} in Sec.~\ref{sec:signal_definition}), is indicated by the black dotted rectangle.
        The concentration of events near $\SI{0.13}{GeV/\textit{c}^{2}}$ in $\megg$ for atmospheric neutrinos originates from single neutral pion production processes.
    }
    \label{fig:all}
\end{figure}

In proton decay occurring in hydrogen atoms in water, the emitted positron and neutral pion are not affected by the intranuclear effects discussed above.
Consequently, the invariant mass reconstructed from the kinematics of the positron and the neutral pion is expected to be consistent with the (free-)proton mass, and the total momentum of the system is expected to be zero.
In contrast, for proton decay events occurring within oxygen nuclei, the intranuclear effects described above, primarily affecting the kinematics of the neutral pion, can lead to substantial deviations of both the reconstructed invariant mass and the total momentum from those of free proton decay.
These two variables, the invariant mass and the total momentum of the system, are used in proton decay searches and provide strong discrimination for proton decay signals against atmospheric neutrino backgrounds.
Figure~\ref{fig:all} shows the two-dimensional distributions of the reconstructed invariant mass, $\megg$, and total momentum, $\pegg$~\footnote{ Throughout this work, we use $\ggonly$ to denote $\pi^0$ and $\egg$ to denote a proton decay candidate: The kinematics are constructed from the individual particles; in the absence of detector or nuclear effects, these are identical to those of the corresponding true particles. }.
Because of nuclear effects, a clear difference in the discrimination power against atmospheric neutrino backgrounds is observed between free and bound proton decay events.

In the following, to illustrate the role of nuclear effects in shaping the event characteristics, we analyzed the distributions of the positron energy ($E_{\positron}$), the $\pi^0$ energy ($E_{\gamma\gamma}$), and the opening angle between their momentum directions ($\thetaegg$)---variables that contain information equivalent to that of the invariant mass and total momentum of the system, and then performed a probabilistic analysis to classify  event candidate as originating from free proton decay, proton decay within oxygen nuclei, or atmospheric neutrino background.

First, for
each of the event categories---free proton decay ($\cfree$), bound proton decay ($\cbound$), or atmospheric neutrino background ($\catmo$)---we constructed a three-dimensional histogram template of $E_{\positron}$ vs. $E_{\gamma\gamma}$ vs. $\thetaegg$. These histograms were generated using \gibuu (see Sec.~\ref{sec:gibuu}) and the detector response of a water Cherenkov detector (see Sec.~\ref{sec:detector_modeling}).

Second, for each test event with a particular $\vec{x} = (E_{\positron}, E_{\gamma\gamma}, \thetaegg)$, we denote the bin content in each template by $H(\vec{x}|\catag)$.

\begin{figure}[!htbp]
    \includegraphics[width=\figwid]{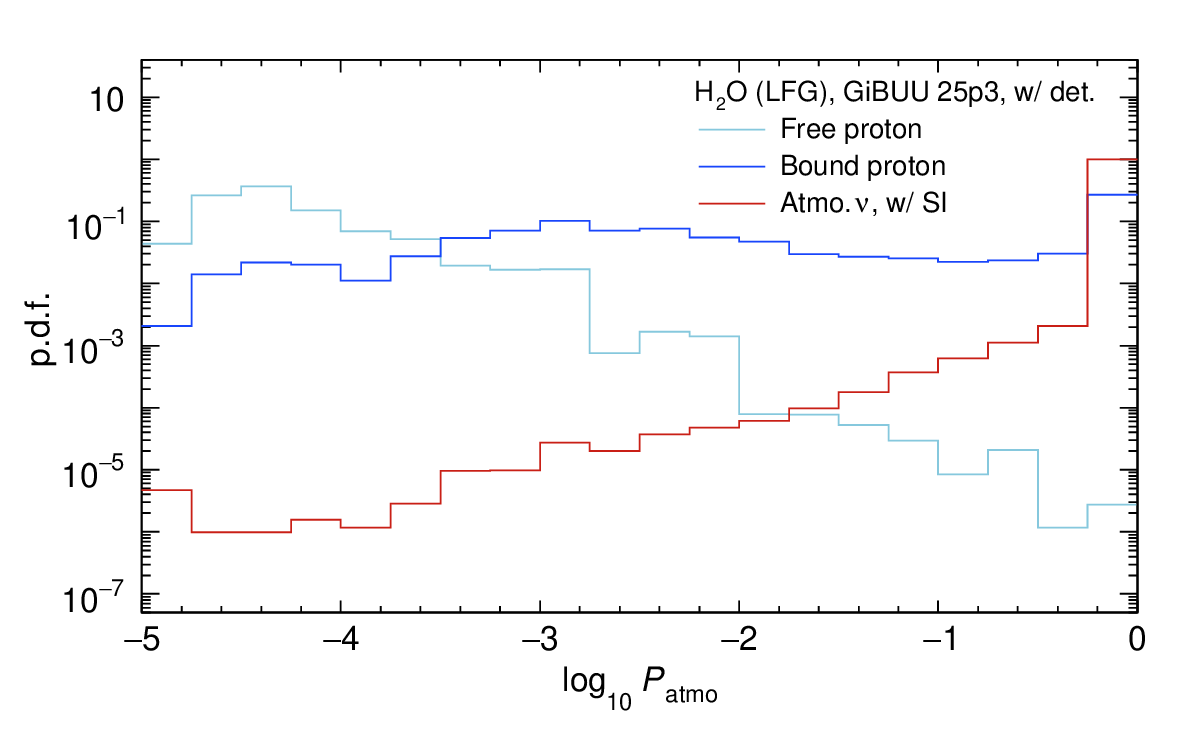}
    \caption{Probability density function (p.d.f.) of  $\log_{10} \patmo$, where $\patmo$ denotes the probability classifier that is defined by
        Eq.~(\ref{eq:categoryprob}).  Underflow events are merged into the first bin.
    }
    \label{fig:posterior_bkg}
\end{figure}

Third, a probability classifier for each category
can be defined as follows:
\begin{align}
     & P_i\equiv \frac{H(\vec{x}|\catag)}{\sum_{j}H(\vec{x}|{\rm C}_{j})},
    \label{eq:categoryprob}
\end{align}
which describes the relative chance of an event originating from $\catag$, inferred from the observed kinematic variables $\vec{x}$.
As an example, Figure~\ref{fig:posterior_bkg} shows the distribution of $\patmo$ for events from different sources.
It can be seen that free proton decays are very efficiently discriminated from the atmospheric background, while a fraction of bound proton decays are misidentified as atmospheric events.

The normalizations of the histogram templates are arbitrary. This is equivalent to choosing \textit{ad hoc} prior probabilities, $\priorp(\catag)$, for each category, such that
\begin{align}
    H(\vec{x}|\catag)=\Bayesp(\vec{x}|{\catag}) \priorp({\catag}),
\end{align}
where $\Bayesp(\vec{x}|{\catag})$ denotes the likelihood. Consequently, $P_i$ corresponds to the Bayesian posterior probability, $\Bayesp({\catag}|\vec{x})$,
\begin{equation}
    P_i=\Bayesp({\catag}|\vec{x}) \equiv \frac{\Bayesp(\vec{x}|{\catag}) \priorp({\catag})}{\sum_j \Bayesp(\vec{x}|{\text{C}_j}) \priorp({\text{C}_j})}.\label{eqn:bayes}
\end{equation}
For data analyses, which are beyond the scope of this work, the discrimination power can be optimized by adjusting the normalization of $H(\vec{x}|\catag)$, i.e.,  by choosing appropriate priors.

\begin{figure}[!htb]
    \includegraphics[width=\linewidth]{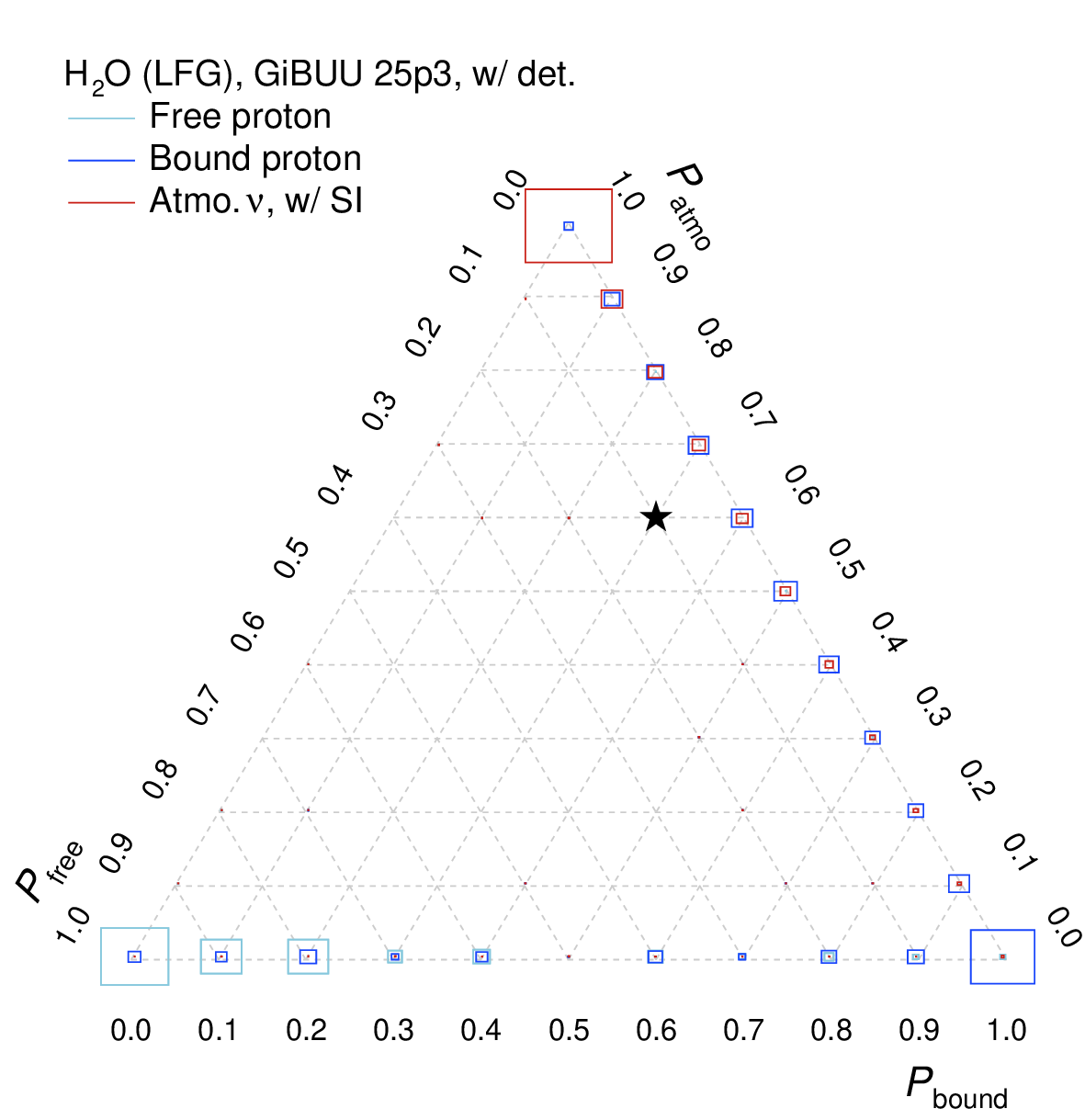}
    \caption{\textbf{Ternary classification of proton decay candidates.}
        Each point inside the triangle corresponds to a unique combination of
        the probability classifiers, $\pfree$, $\pbound$, and $\patmo$, defined by Eq.~(\ref{eq:categoryprob}).
        As an example, the star marker  indicates $\pfree = 0.1$, $\pbound = 0.3$, and $\patmo = 0.6$.
        The size of each box is proportional to the
        event population, shown on a linear scale.
        Individual plots for each event type are shown in Appendix~\ref{app:dec}.
    }
    \label{fig:3sample}
\end{figure}

Regardless of the normalization of the templates, it is evident from Eq.~(\ref{eq:categoryprob}) that
\begin{align}
    \pfree+\pbound+\patmo=1.
\end{align}
A classification can therefore be represented conveniently in a ternary plot, as shown in Fig.~\ref{fig:3sample}.
All events shown with red boxes are true atmospheric neutrino background events: some are correctly classified as background, while others are misclassified as bound proton decay signal, but few are classified as free proton decay signal.
In contrast, true free proton decay events (light-blue boxes) are classified almost exclusively as proton decay signals and not as atmospheric background. Such a hierarchical classification is the result of the good separation between free proton decay and background events observed in \figref{fig:posterior_bkg}.  It demonstrates that nuclear effects on the signal can make it more background‑like, thereby increasing the difficulty of distinguishing signal from background.

\begin{figure}[!htb]
    \includegraphics[width=\linewidth]{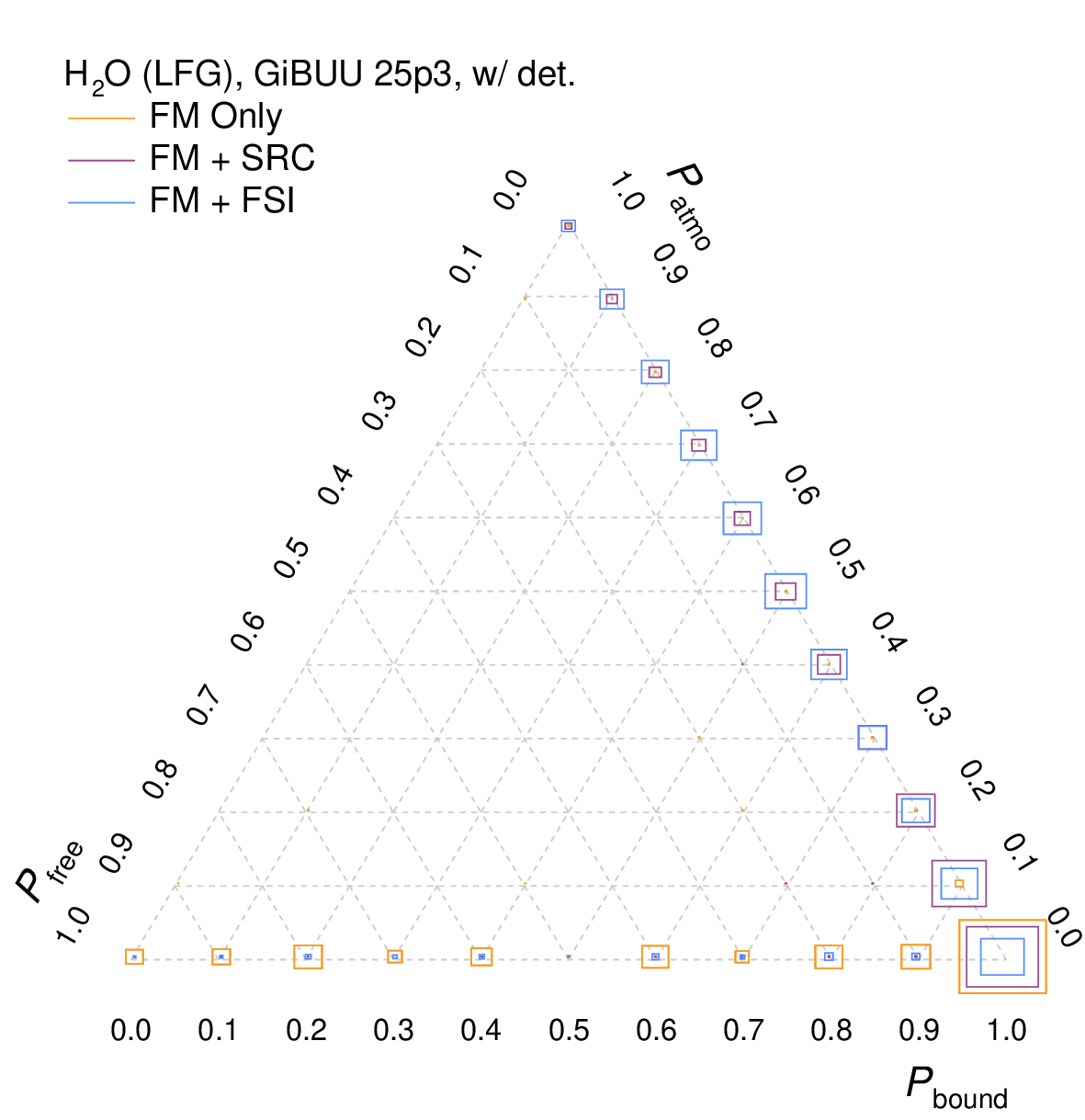}
    \caption{Similar to \figref{fig:3sample}, but restricted to true bound proton decay events. In the legends, FM, SRC, and FSI denote Fermi motion, short-range correlation, and final-state interaction, respectively.
    }
    \label{fig:split_oxygen}
\end{figure}

The impact of the nuclear effects is further illustrated in \figref{fig:split_oxygen}, where the bound proton decay signal is subdivided according to the presence or absence of FSI and SRC effects.
For event samples without SRCs or FSIs, the separation from the background is  pronounced, with mixing occurring mainly between free and bound proton decay signals.
When these effects are included, bound proton decay events are more likely classified as atmospheric neutrino background.

\section{The \gibuu model}\label{sec:gibuu}

\gibuu is a comprehensive
model designed to simulate a wide range of nuclear reactions, including those induced by neutrinos, electrons, photons, and hadrons~\cite{Buss:2011mx}.
It is based on the Boltzmann-Uehling-Uhlenbeck (BUU) equation, which describes the time evolution of the phase-space distribution functions of particles within a nuclear medium. \gibuu treats the entire reaction process in a unified framework, from the initial interaction to the final state particles exiting the nucleus. Being widely used and tested extensively in a range of research areas, including nuclear reactions and heavy-ion collisions~\cite{Gallmeister:2009ht,Larionov:2020fnu, Lehr:2003ht, Buss:2006vh, Yan:2025aau, Bogart:2024gmb, Gallmeister:2016dnq,HADES:2023sre}, \gibuu has been validated against a variety of experimental data, demonstrating its capability to accurately model complex nuclear processes.

For a mean-field description of the nucleus, we use the Thomas-Fermi  LFG implemented in \gibuu~\cite{Buss:2011mx,Mosel:2023zek}, which provides a realistic representation of the nucleon momentum distribution, including the effects of FM and binding energy but without SRC.
Alternatively, \cda, rather than the correlation-function approach of Ref.~\cite{Yamazaki:1999gz},
is used to account for the high-momentum tails of SRCs.

A key difference between \gibuu and INC-based approaches, including those implemented in generators, lies in the treatment of the nuclear medium and particle propagation.
While INC models typically treat nucleons as free particles within a potential well and simulate interactions as a series of independent collisions, \gibuu incorporates mean-field potentials that account for the collective behavior of hadrons in the nucleus. This allows \gibuu to capture important nuclear effects such as the Fermi motion and binding energy between nucleons more accurately and consistently \cite{Gallmeister:2025jwe}.

In \gibuu, an additional phenomenon---the broadening of the $\Delta$ resonance width at the primary interaction vertex of, e.g., an atmospheric background---can be considered using the Oset model~\cite{Oset:1987re,Salcedo:1987md}, which accounts for both Pauli blocking and collisional broadening. The former results in a reduced $\Delta$ width due to a limited decay phase space, while the latter arises from the $\Delta$ interactions with the medium.
The net effect is a broadening of the $\Delta$ width, which is more pronounced at higher nuclear densities. In \gibuu, the imaginary part of the $\Delta$ self-energy, as parametrized by Oset \textit{et al.}~\cite{Oset:1987re}, is implemented. For the real part of the $\Delta$ self-energy no such parametrization is available. In \gibuu, we, therefore, work with a $\Delta$ potential linked to the nucleon potential $U_{\rm N}$ by $U_\Delta = 2/3\,  U_{\rm N}$ following results obtained from the $\Delta$-hole model~\cite{Salcedo:1987md,Ericson:1988gk}.

The broadening of the $\Delta$ leads to
lower $\pi$ production in neutrino-nucleus interactions~\cite{Yan:2025aau}, thereby affecting the background event rate in proton decay searches.
This so-called ``Oset broadening" also applies to the cross section of $\pi\nucleon$ scattering during FSIs in an atmospheric background or following a proton decay. It can affect the absorption, elastic scattering, and charge exchange of pions in the nucleus, and therefore
the topology and kinematics of the final state particles.

Another effect, the so-called ``in-medium modification", introduces a medium density dependent suppression of the nucleon-nucleon ($\nucleon\nucleon$) elastic scattering cross section~\cite{Li:1993rwa,Li:1993ef} or $\pi$ inelastic scattering cross section~\cite{Song:2015hua}.
The suppression of the $\pi$ inelastic scattering cross section can be expressed in the following medium-density-dependent form:
\begin{equation}   \label{eq:Delta-supp}
    \sigma_{\nucleon\nucleon\rightarrow \nucleon\Delta}(\rho_\nucleon)  = \sigma_{\nucleon\nucleon\rightarrow \nucleon\Delta}(0) \exp\left\{-\alpha \frac{\rho_\nucleon}{\rho_0}\right\},
\end{equation}
where $\rho_\nucleon$ is the local nuclear density at the interaction point, and $\alpha$ is a free parameter.
Because of time-reversal invariance, this density-dependence also affects the inverse cross section for the pionless decay of the Delta ($\Delta\nucleon \to \nucleon\nucleon$)  and thus the reabsorption of pions.

\section{Proton decay sensitivity in future experiments}\label{sec:setup-and-analysis}

In this work, we established a comprehensive simulation and analysis framework to evaluate the sensitivity of future proton decay searches in the $\pdk$ channel.
This framework integrates detailed modeling of proton decay events, atmospheric neutrino background events, and the detector response, including typical event reconstruction effects in modern water Cherenkov detectors.
Systematic uncertainties associated with nuclear effects were thoroughly assessed using the \gibuu model.

\subsection{Proton decay event generation}\label{sec:eventgen_pdk}
Proton decay is assumed to occur with uniform probability for protons in water molecules. Decays of hydrogen atoms (free protons) were generated by producing a positron and a neutral pion emitted in a back-to-back configuration.

\begin{figure}[!htbp]
    \centering
    \begin{subfigure}{\figwid}
        \centering
        \includegraphics[width=\figwid]{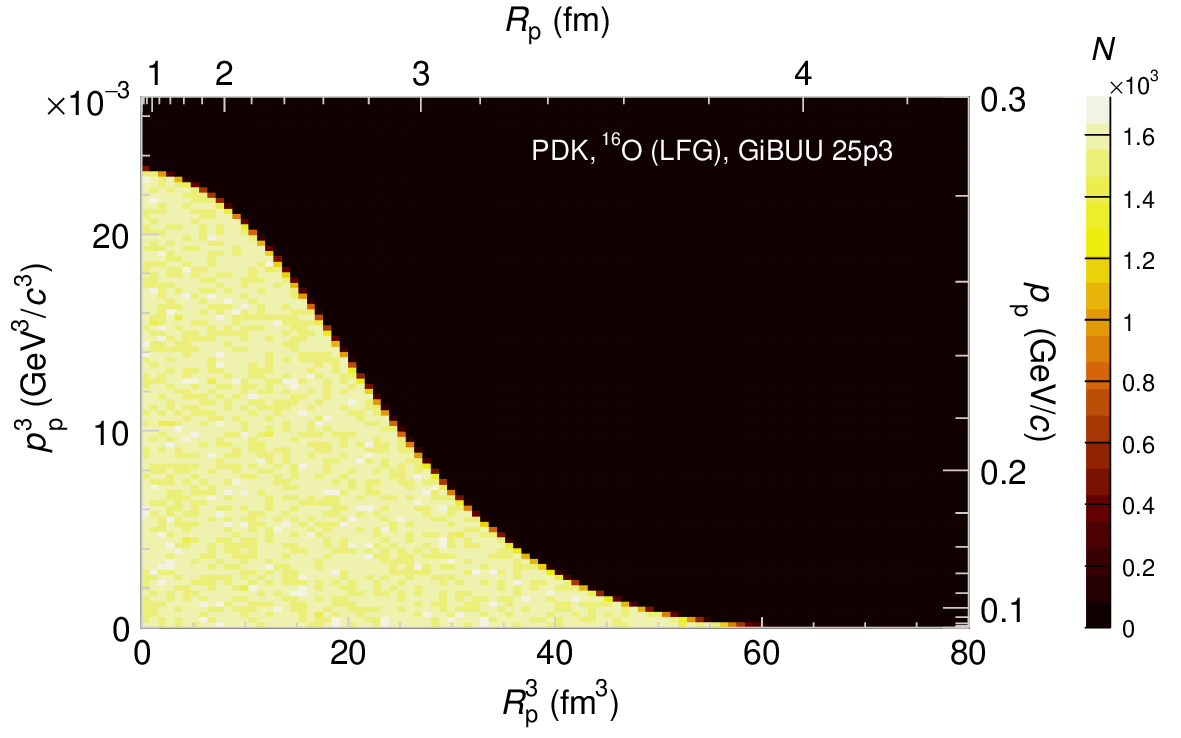}
        \caption{Phase space distribution.
        }\label{fig:r3p3a}
    \end{subfigure}
    \begin{subfigure}{\figwid}
        \centering
        \includegraphics[width=\figwid]{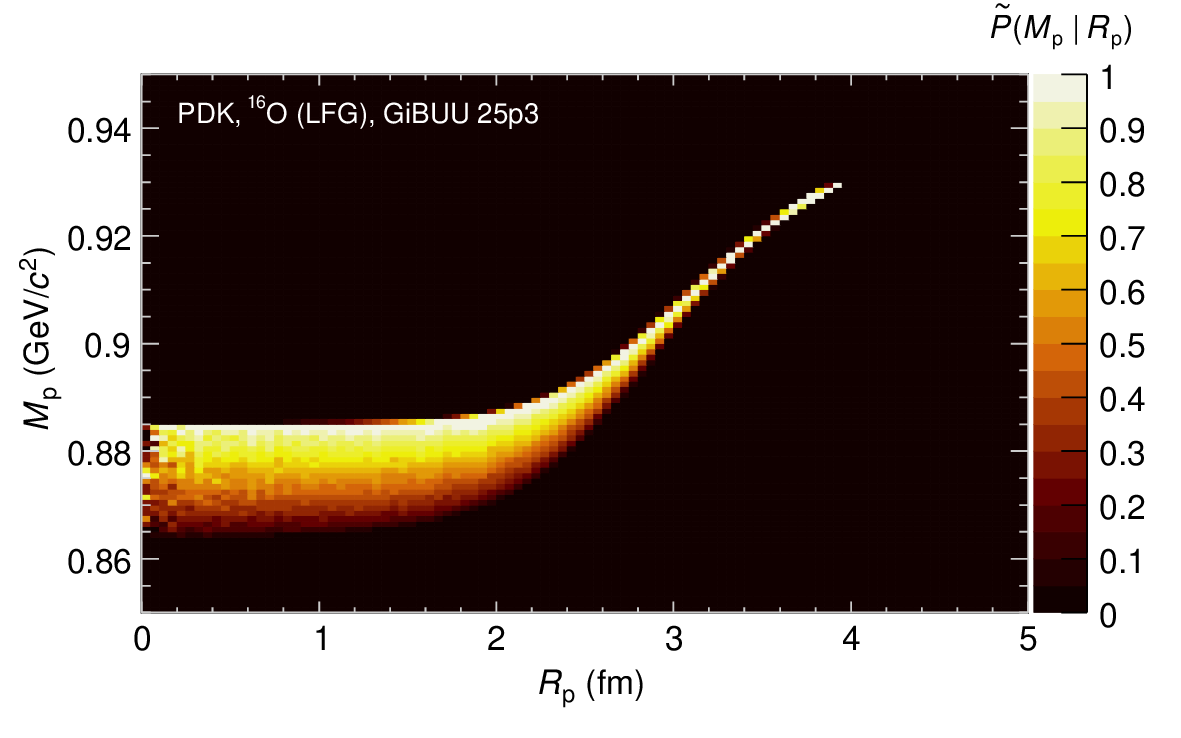}
        \caption{
            Conditional probability $\tilde{P}\left(M_\proton|R_\proton\right)$ renormalized to the peak value, i.e., the maximum of each $R_\proton$-slice is 1.
        }\label{fig:rMa}
    \end{subfigure}
    \begin{subfigure}{\figwid}
        \centering
        \includegraphics[width=\figwid]{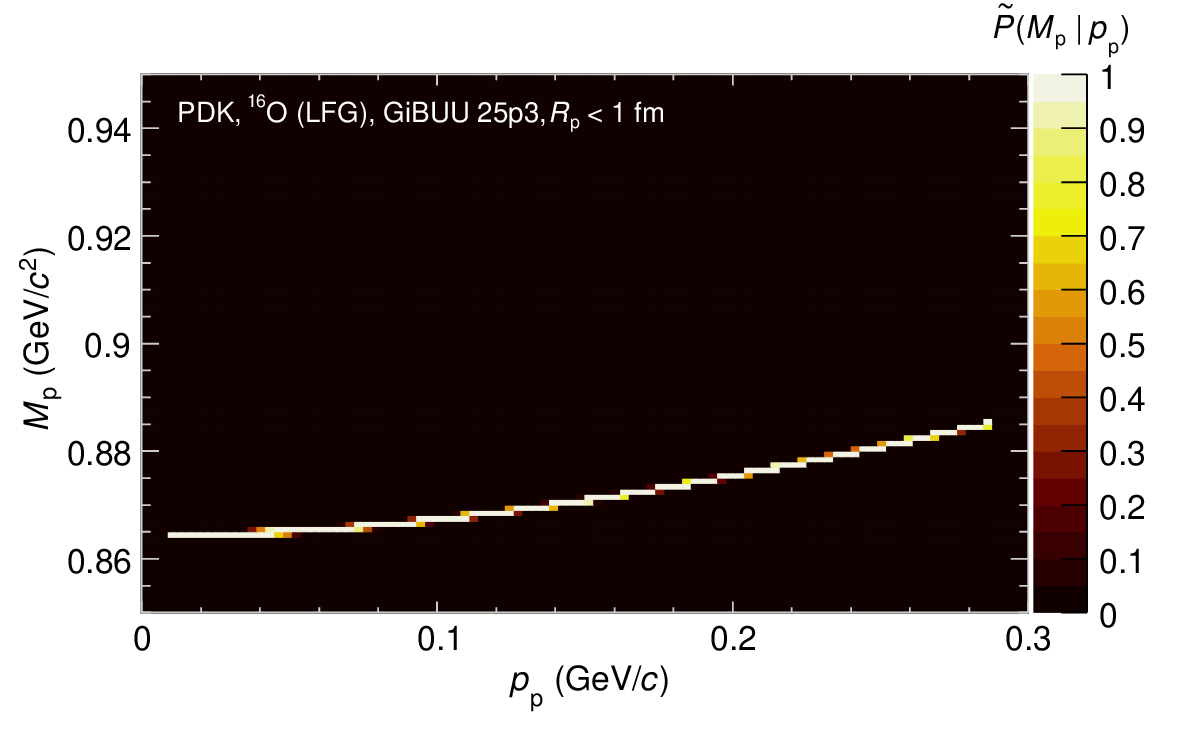}
        \caption{%
            $\tilde{P}\left(M_\proton|p_\proton\right)$ for  $R_\proton<1~\textrm{fm}$.
        }\label{fig:lfg_p3_ma}
    \end{subfigure}
    \caption{
        The initial proton location ($R_{\proton}$, the radial distance from the nucleus center), momentum ($p_{\proton}$), and invariant mass ($M_{\proton}\equiv\sqrt{E_\proton^2-p_\proton^2}$ where $E_\proton$ is the simulated proton energy) in the sampled proton decay (PDK) events. The nuclear initial state is LFG.  Figures for the Ciofi degli Atti–Simula (\cda) model~\cite{CiofidegliAtti:1995qe} can be found in Appendix~\ref{app:initial_state}. As a comparison,  see also Fig.~1 of Ref.~\cite{Yan:2025aau} for the mass and momentum evolution during FSIs in neutrino interactions.
    }\label{fig:inirpma}
\end{figure}

The simulation of proton decay events within the oxygen nucleus (bound protons) was performed using \gibuu (version \gibver).
The initial-state proton was sampled from the LFG model of oxygen:
the model assigns a nucleon position and momentum corresponding to the local nuclear density; an initial proton momentum distribution is generated with a constant probability within the local Fermi sphere, filling the allowed phase space up to the Fermi surface.
As is shown in Fig.~\ref{fig:r3p3a}, the Fermi surface goes up to about 280~MeV$/c$ at the nuclear center due to the high density, and decreases rapidly to below 200~MeV$/c$ at a radial location, $R_\proton$, beyond 3~fm.
Because of the mean-field potential, the proton is off-shell, with the largest deviation---up to about 50~MeV$/c^2$ below the bare mass---occurring at the nuclear center, where the binding is strongest (Fig.~\ref{fig:rMa}).
By restricting the radial location, e.g., $R_\proton<1$~fm (Fig.~\ref{fig:lfg_p3_ma}), we can see that the off-shellness is stronger at lower momentum, indicating even stronger binding~\cite{Buss:2011mx}.

\begin{figure}[htbp]
    \includegraphics[width=\figwid]{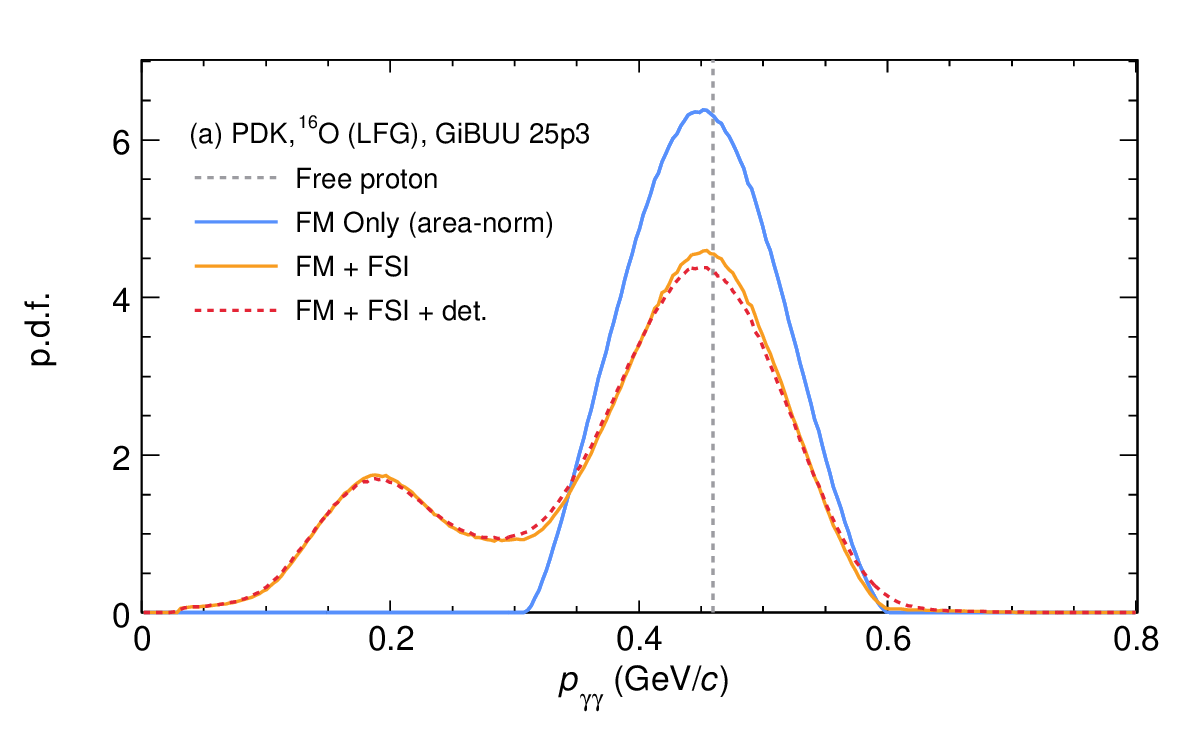}
    \includegraphics[width=\figwid]{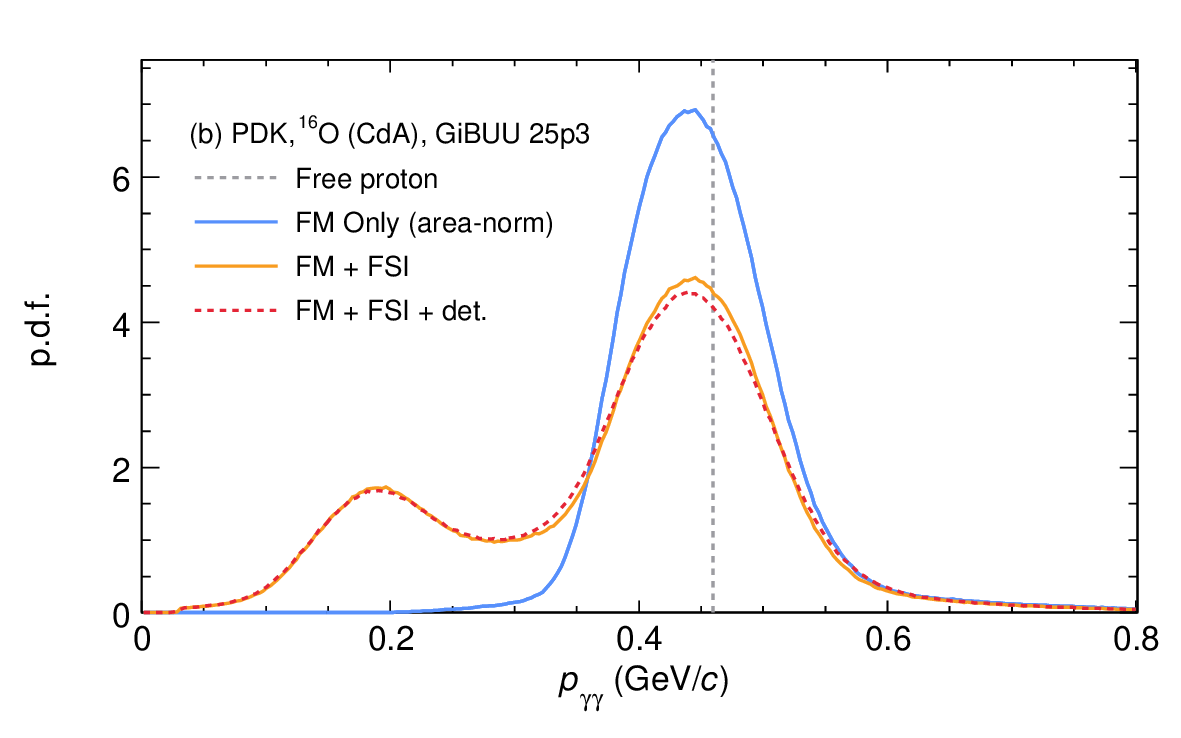}
    \caption{
        $\pi^0$ momentum ($\pgg$)  distribution from free- and bound-proton decays comparing (a) LFG and (b) \cda.
        The ``FM only'' distribution is area-normalized, while the others are physically normalized to it.
    }
    \label{fig:pi0p}
\end{figure}

In \gibuu the nucleon mean field  is given by a Skyrme-potential in the local rest frame. To simplify the transformation to a moving frame, the potential is then assumed to be of a Lorentz-scalar nature, thus effectively lowering the mass of the nucleon in the medium (for a detailed discussion see Sect. 3.1 in \cite{Buss:2011mx}). The decaying proton's initial mass ($M_\proton$) is then given by: 
\begin{equation}
	M_\proton^2 = E_\proton^2 - \left|\vec{p}_\proton\right|^{2} = (M_\text{free} - U)^2 ,
\end{equation}
where $U$ is the positive, position- and momentum-dependent mean-field potential, $\vec{p}_\proton$ is the sampled initial proton momentum, and $M_\text{free}$ is the bare proton mass. This reduction in the invariant mass shifts the $\pi^0$ momentum distribution toward lower values compared to the free proton decay case, as shown in \figref{fig:pi0p}.
Prior to the inclusion of FSI effects, the width of the $p_\ggonly$ distribution is driven by the initial momentum of nucleons in the oxygen nucleus, while the peak is shifted to lower momentum relative to the free-proton case due to nuclear binding energy.

The products of the proton decay,
$\pdk$, were generated isotropically in the rest frame of the initial proton and then boosted to the laboratory frame using the sampled initial proton momentum. The hadronic final state, i.e., $\pi^0$ in this case, was subsequently propagated through the nuclear medium using \gibuu.
FSIs can remove momentum from the $\pi^{0}$, creating a lower-momentum peak in the momentum distribution as shown in \figref{fig:pi0p}.

\begin{figure}[htbp]
    \includegraphics[width=\figwid]{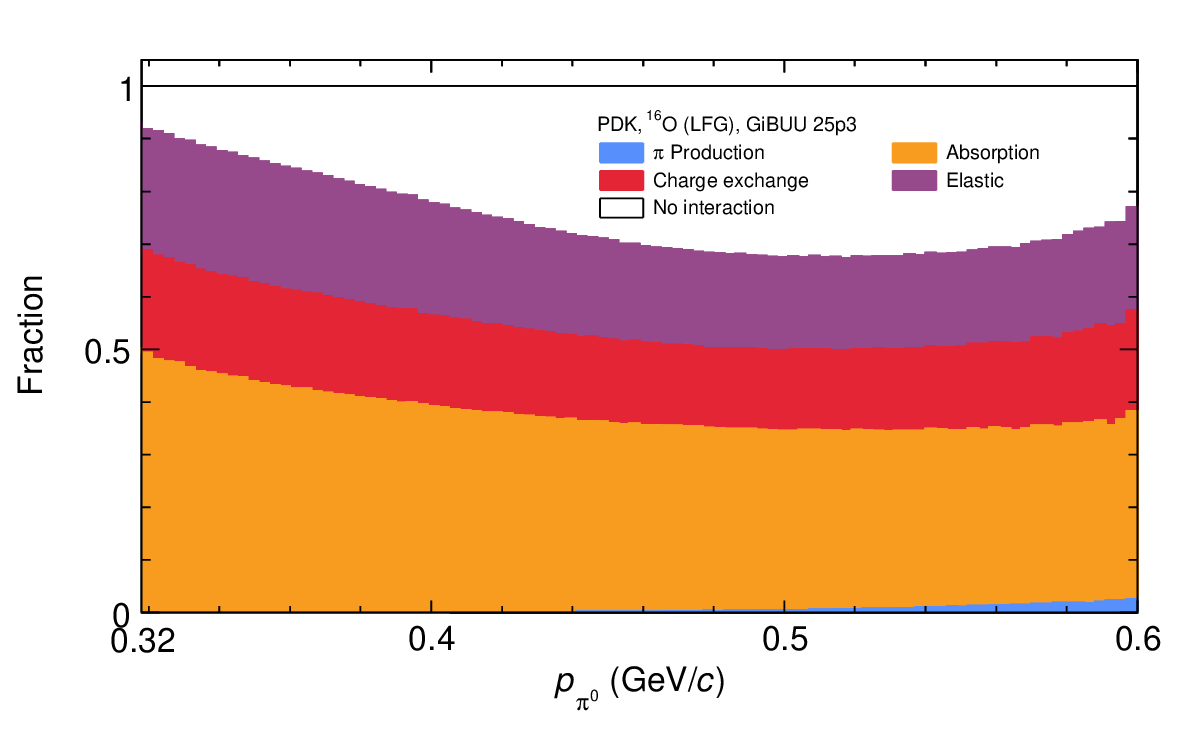}
    \caption{
        Fractions of $\pi^{0}$'s undergoing FSIs, as a function of the
        pion momentum at production, $p_{\pi^0}$.
        In the legends, ``no interaction'' refers to events where the final-state $\pi^0$ shows no change in its momentum; ``elastic'' refers to final states with no extra particles produced and no existing particles absorbed; ``charge exchange'' refers to final states where a $\pi^0$ changes its charge via interaction with nucleons, producing a $\pi^\pm$; ``absorption'' refers to events where the $\pi^0$ is absorbed, leaving no pion in the final state; and ``$\pi$ production'' refers to processes where new pions are produced, resulting in more than one pion in the final state.
        The range of the $\pi^0$ momentum at production is determined by Fermi motion, as illustrated by the ``FM Only'' curve in Fig.~\ref{fig:pi0p}a.
    }
    \label{fig:pi0fate}
\end{figure}

For events in which the $\pi^0$ undergoes inelastic scattering or absorption, the final state may not contain a $\pi^0$ at all, resulting in an unavoidable loss in detection efficiency.
The fate of the $\pi^{0}$ during FSIs is illustrated in \figref{fig:pi0fate}: approximately 40\% of the pions are absorbed when their momentum at production lies between 0.4 and 0.5~GeV/$c$.

After exiting the nucleus, the $\pi^0$ decays almost instantaneously into two photons ($\pi^0 \to 2\gamma$).
The momenta and directions of the two photons were sampled isotropically in the $\pi^0$ rest frame and then boosted to the laboratory frame. The four‑momenta of all final-state particles, namely, the positron and the two photons, were recorded for subsequent analysis.

\subsection{Atmospheric neutrino background event generation}\label{sec:eventgen_atm}
The atmospheric neutrino background originates from interactions of $\nu_e$, $\nu_\mu$, $\bar{\nu}_e$, and $\bar{\nu}_\mu$ with nuclei in water molecules, including both charged-current (CC) and neutral-current (NC) processes. The neutrino flux used as input is the Honda flux~\cite{Honda:2015fha} for the Super-Kamiokande site. Neutrino-nucleus interactions were simulated using the same \gibuu setting as for the proton decay event generation.

Even when no neutral pions or neutrons---handles for atmospheric neutrino background rejection---are produced at the primary neutrino interaction vertex, hadrons emitted after the primary neutrino interaction can subsequently generate neutral pions or neutrons through secondary hadronic interactions in water.
Such processes can affect proton decay search analyses.
To account for these secondary interactions (SIs), we incorporate their simulation using the publicly available \geant-based (version 10.3.3~\cite{GEANT4:2002zbu, Allison:2006ve, Allison:2016lfl}) water Cherenkov detector simulation package, \wcsim~\cite{WCSim}.
Hadronic interactions in the energy range relevant to this study are modeled using the Bertini intranuclear cascade model~\cite{Wright:2015xia}, while neutron transport below 20~MeV is treated with the high-precision (HP) data-driven model based on evaluated nuclear data libraries~\cite{IAEA_NDS_0758}.
The de-excitation of excited nuclei is described by the \geant precompound model~\cite{precompound}.
Note that, the $\pi^0$ from the proton decay will decay immediately into two photons before any SI happens, and therefore, SIs are only considered for atmospheric neutrino interaction backgrounds.

\begin{figure}[!htbp]
    \centering
    \includegraphics[width=\figwid]{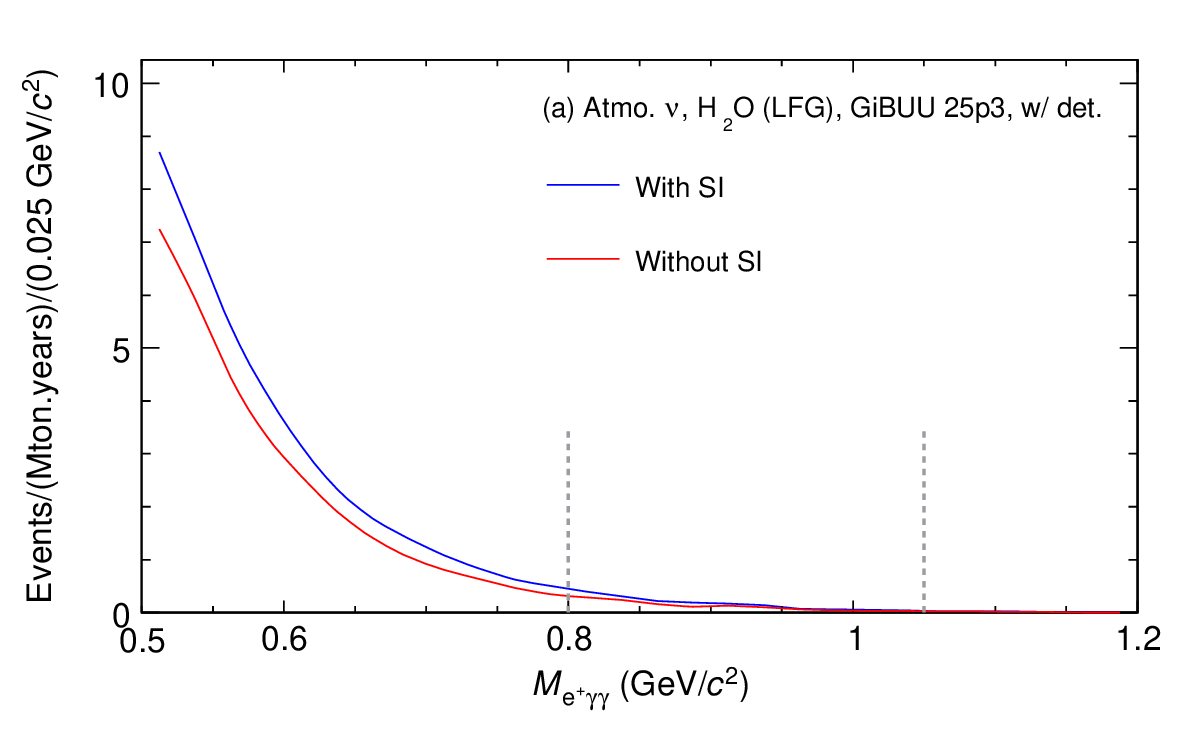}
    \includegraphics[width=\figwid]{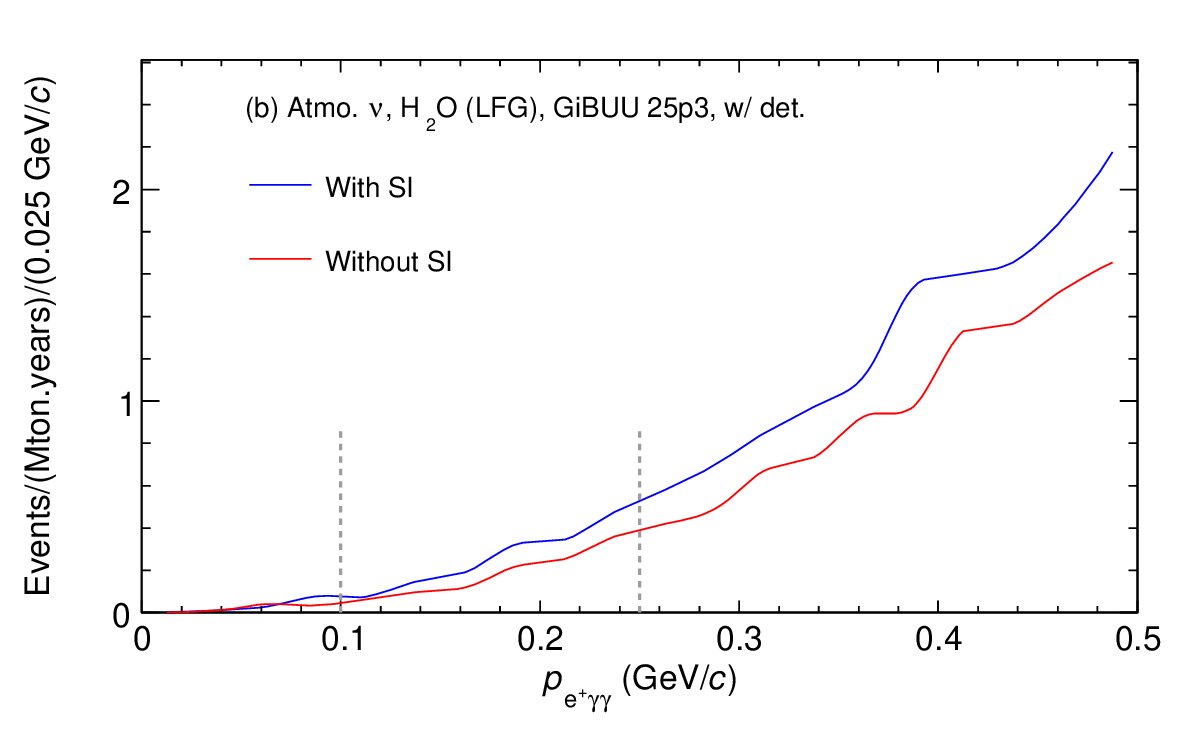}
    \caption{
        (a) $M_\egg$ and (b) $p_\egg$ from atmospheric neutrino background events with detector smearing, comparing the cases with (blue curve) and without (red curve) secondary interactions in the detector simulation.
        For (a), all selection criteria up to \textbf{C6} (see Sec.~\ref{sec:signal_definition}) except \textbf{C5} are applied; for (b), all selection criteria up to \textbf{C5} are applied.
    }\label{fig:C6_total_m_limit}
\end{figure}

During the simulation of secondary interactions from atmospheric neutrinos, we track the production of neutral pions and the capture of neutrons.
Events in which charged pions undergo charge exchange to produce a $\pi^0$ are treated as if the $\pi^0$ originated from the primary interaction vertex.
And the extra neutral pions produced via secondary interactions contribute to the background contamination in the proton decay search in addition to those produced at the primary interaction vertex.
The impact of charge exchange during secondary interactions on the final-state total momentum and invariant mass distributions is shown in~\figref{fig:C6_total_m_limit}, demonstrating a significant enhancement of atmospheric neutrino background events  in the kinematic range of events that pass the selection criteria
(see Sec.~\ref{sec:signal_definition}).

\begin{figure}[!htbp]
    \includegraphics[width=\figwid]{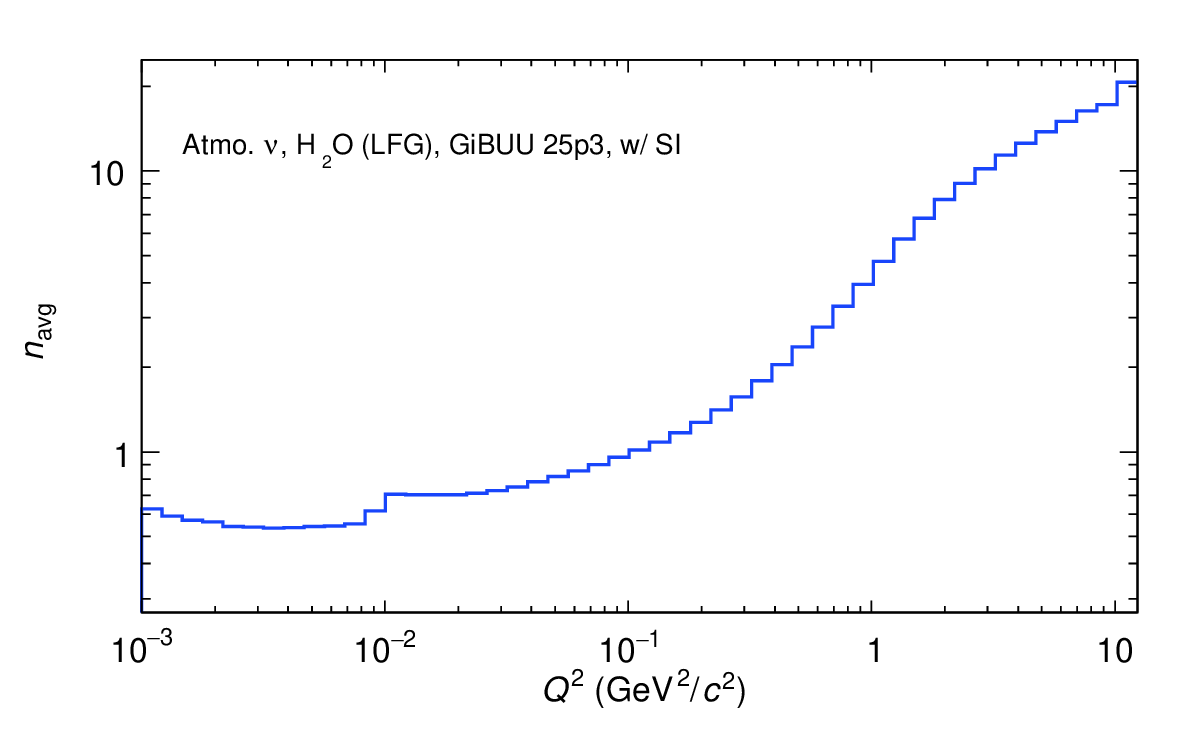}
    \caption{MC truth correlation between average number of captured neutrons, $n_\textrm{avg}$, and the squared four-momentum transfer, $Q^2$, for atmospheric neutrino interactions in water.
    }
    \label{fig:si_ncapture}
\end{figure}

In our simulation, the number of captured neutrons is recorded, and a realistic neutron tagging efficiency in future water Cherenkov detectors is applied to assess the impact of neutron tagging on background rejection (see Sec.~\ref{sec:signal_definition}).
Figure~\ref{fig:si_ncapture} shows the relationship between the average number of neutrons captured in water, including those produced via secondary interactions following the neutrino interaction, and the squared four-momentum transfer to the nucleon, $Q^{2}$, at the neutrino interaction vertex.
As the momentum transfer to the hadronic system increases, the number of produced neutrons also increases.
This trend is consistent with that observed in atmospheric neutrino events in the Super-Kamiokande experiment~\cite{Super-Kamiokande:2025cht}.
As noted in the same reference, the average number of neutrons produced in water can vary by up to approximately $\pm$20\% depending on the choice of hadronic interaction model.
Since evaluating the impacts on the remaining backgrounds due to this variation in neutron yield is beyond the scope of the present study, it is not considered here, but it remains an interesting subject for future investigations.

\subsection{Detector response modeling}\label{sec:detector_modeling}

The detector response and reconstruction effects were modeled by applying smearing to the momenta and directions of the final-state particles. This smearing accounts for two key aspects: angular resolution and momentum resolution.

Angular resolution was modeled using a distribution of the opening angle between the measured and true particle directions, taken from Fig.~6.16 of Ref.~\cite{Tobayama:2016dsi}: The distribution of the angular difference between the reconstructed and true electron directions is well described by a Rayleigh form; for an electron with a momentum of $\SI{500}{MeV/\textit{c}}$, the 68\% containment angle is $2.81^\circ$.
The momentum dependence of the angular resolution was incorporated according to Fig.~6.17 of the same reference.
The reference provides angular resolutions only for electrons and muons in single-particle (single Cherenkov ring) events, while multiple particles (Cherenkov rings), including high energy photons from the $\pi^{0}$ decay, are expected to be observed in proton decay events.
Therefore, we introduced a per-particle-type scaling factor to extend the model to other particle types and multi-particle event cases.
This scaling factor is applied directly to the resolution values extracted from the reference, thereby adjusting for differences in detection characteristics among the various particle species.

The same approach was used to model the momentum resolution. Gaussian distributions centered on the true momentum values were used for smearing, with the widths of these distributions (i.e., the momentum resolutions) extracted from Fig.~6.18 of Ref.~\cite{Tobayama:2016dsi} for single-particle events: the momentum resolution for electrons is approximately $3.5\%$ at a momentum of $\SI{500}{MeV/\textit{c}}$.
For the same reason as for the angular resolution, a per-particle-type scaling factor was introduced to extend the momentum resolution model to other particle types and to better represent the characteristics of proton-decay events.

These free scaling parameters (two per particle type)---controlling the particle direction and momentum resolutions---were tuned to reproduce the invariant mass distributions of the $\pi^0$ and free proton decay events observed in the Super-Kamiokande simulation~\cite{Takenaka:2020cmb}.
The tuning strategy proceeded as follows. First, the angular and momentum resolutions for photons were adjusted to match the invariant mass distribution of the outgoing $\pi^0$ (reconstructed from the two photons).
Next, the resolutions for electrons/positrons were tuned to reproduce the invariant mass distribution of the $\positron\pi^0$ system in free proton decay events. The final scaling parameters obtained from this tuning process were then applied consistently to all particle types in the free and bound proton decay signal and atmospheric neutrino background simulations.
Note that this tuning does not make use of event samples involving nuclear effects; therefore, differences in the nuclear interaction models between this work and the Super-Kamiokande analysis do not affect the result.

The Super-Kamiokande reconstruction algorithm captures Cherenkov rings projected onto the photosensors mounted on the detector wall to identify the number of particles in observed events.
This Cherenkov ring counting capability was accounted for according to Ref.~\cite{Shiozawa:1999sd}, by assuming that the detector can separate particles with an opening angle greater than $15^\circ$ between their particle directions.
Particles with an opening angle between them less than $15^\circ$ were merged into a single particle, with their momenta summed. After merging, any particle with momentum less than $\SI{30}{MeV/\textit{c}}$ was considered undetectable and removed from the event.

The separation capability between showering (e-like) and non-showering ($\mu$-like) particles in the Super-Kamiokande detector is estimated to be very high, with an efficiency of approximately 99\%.
Therefore, in this mock detector, the particle identification (PID) for showering and non-showering particles is assumed to be perfect.
The efficiency for tagging Michel electrons from muon decay is assumed to be 88\%~\cite{Super-Kamiokande:2017yvm}.

In this study, a mock detector response is introduced based on the event reconstruction performance of the Super-Kamiokande detector as described above.
These reconstruction resolutions depend primarily on the number of detected Cherenkov photons.
The Hyper-Kamiokande detector is expected to employ photomultiplier tubes with approximately twice the photon detection efficiency of those used in Super-Kamiokande, while its photocathode coverage is planned to be about half that of Super-Kamiokande~\cite{Hyper-Kamiokande:2025fci}.
As a result, the overall number of detected photons per unit particle energy is expected to be comparable to that of the Super-Kamiokande detector, and no significant difference in event reconstruction performance is therefore anticipated.

\begin{figure}[!htbp]
    \includegraphics[width=\figwid]{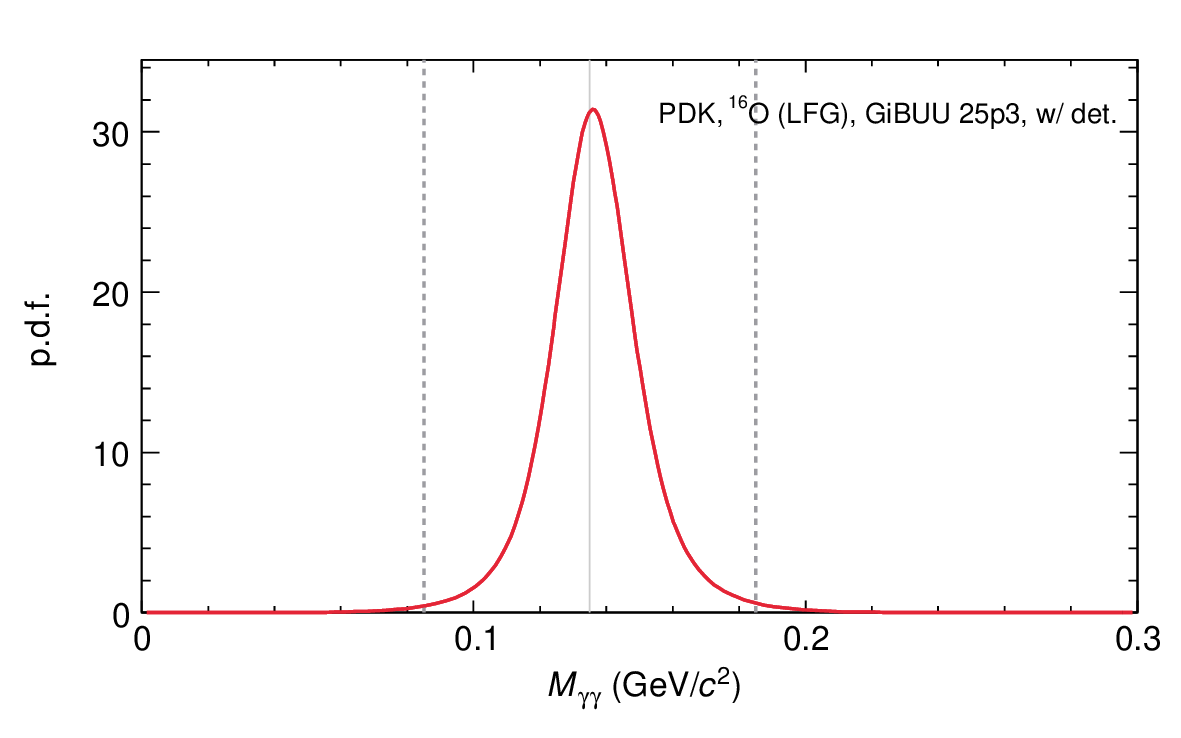}
    \caption{Reconstructed $\pi^0$ invariant mass, $M_\ggonly$, from proton decay events.
        The true $\pi^0$ mass is indicated by the solid gray line.  The signal definition cut, \textbf{C4}, is indicated by the two vertical dashed lines.
    }
    \label{fig:pi0M}
\end{figure}

The detector response smears the kinematic distributions of the two photons from the $\pi^0$ decay, as shown in \figref{fig:pi0M}. The reconstructed $\pi^0$ invariant mass distribution peaks near the nominal $\pi^0$ mass of \SI{135}{MeV/\textit{c}^2}, but exhibits significant broadening due to detector effects.
Figure~\ref{fig:pi0p} also shows additional smearing to the distribution of $\pi^0$ momentum, $\pgg$, by reconstruction effects.

\begin{figure}[!htbp]
    \includegraphics[width=\figwid]{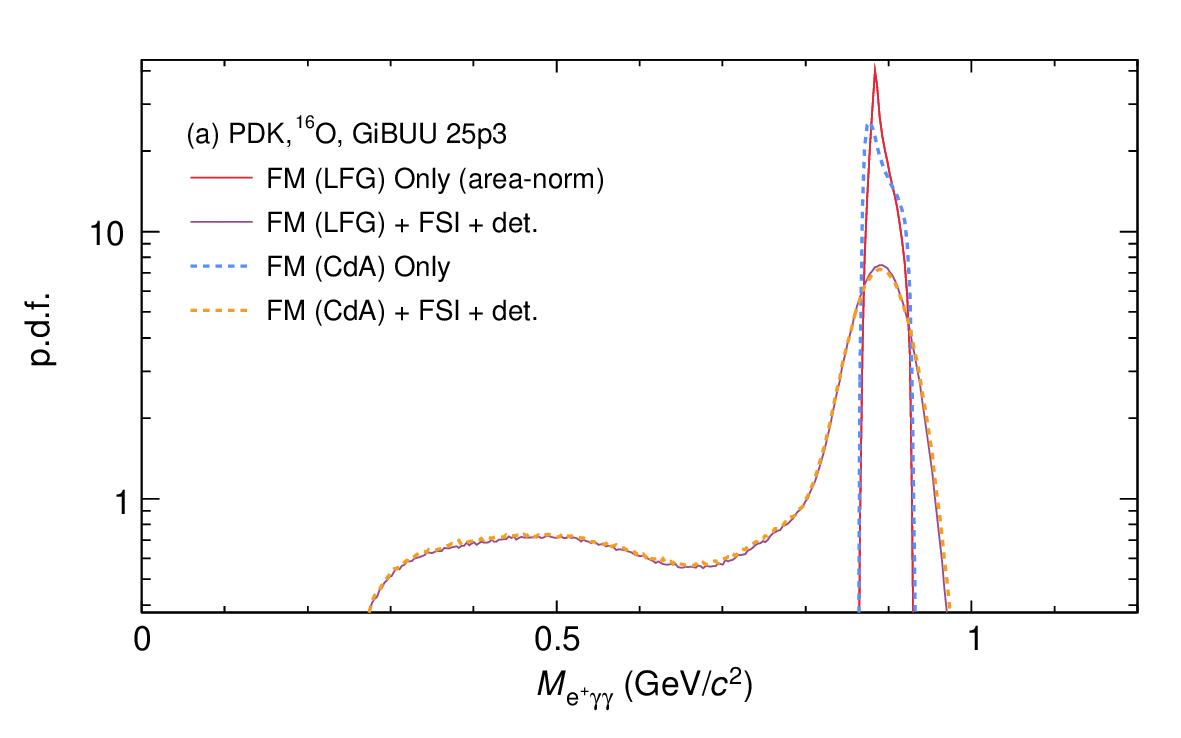}
    \hfill
    \includegraphics[width=\figwid]{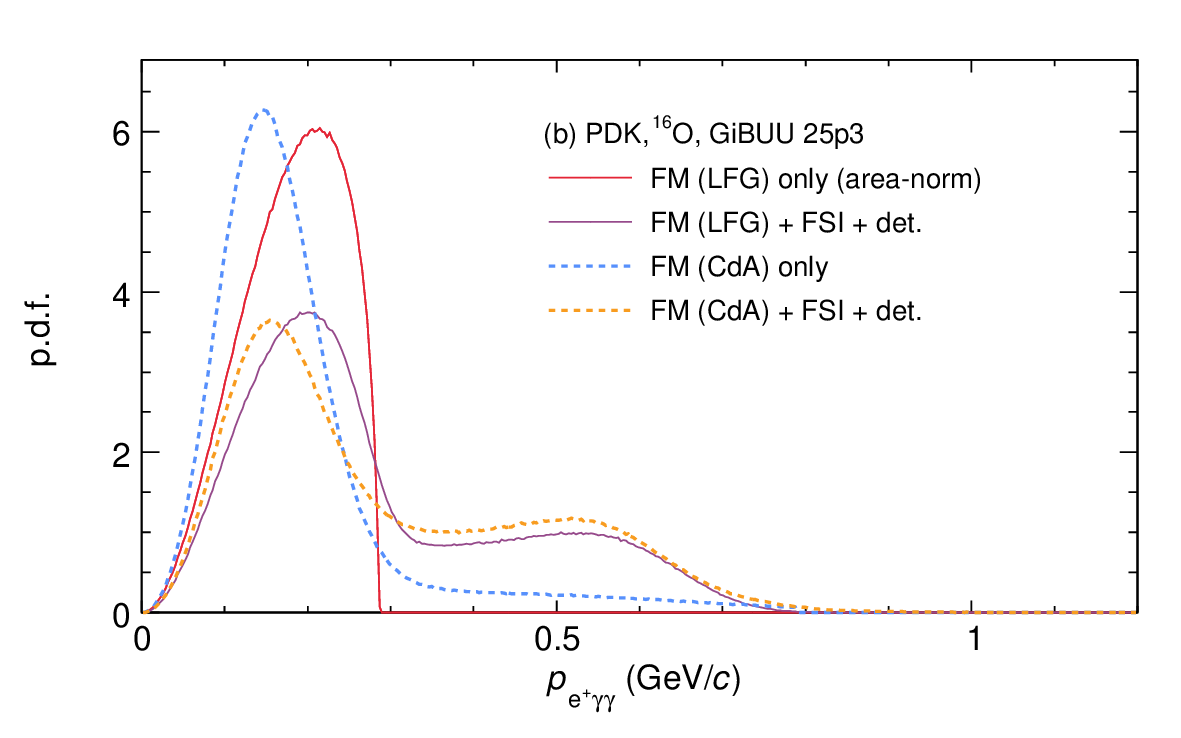}
    \hfill
    \includegraphics[width=\figwid]{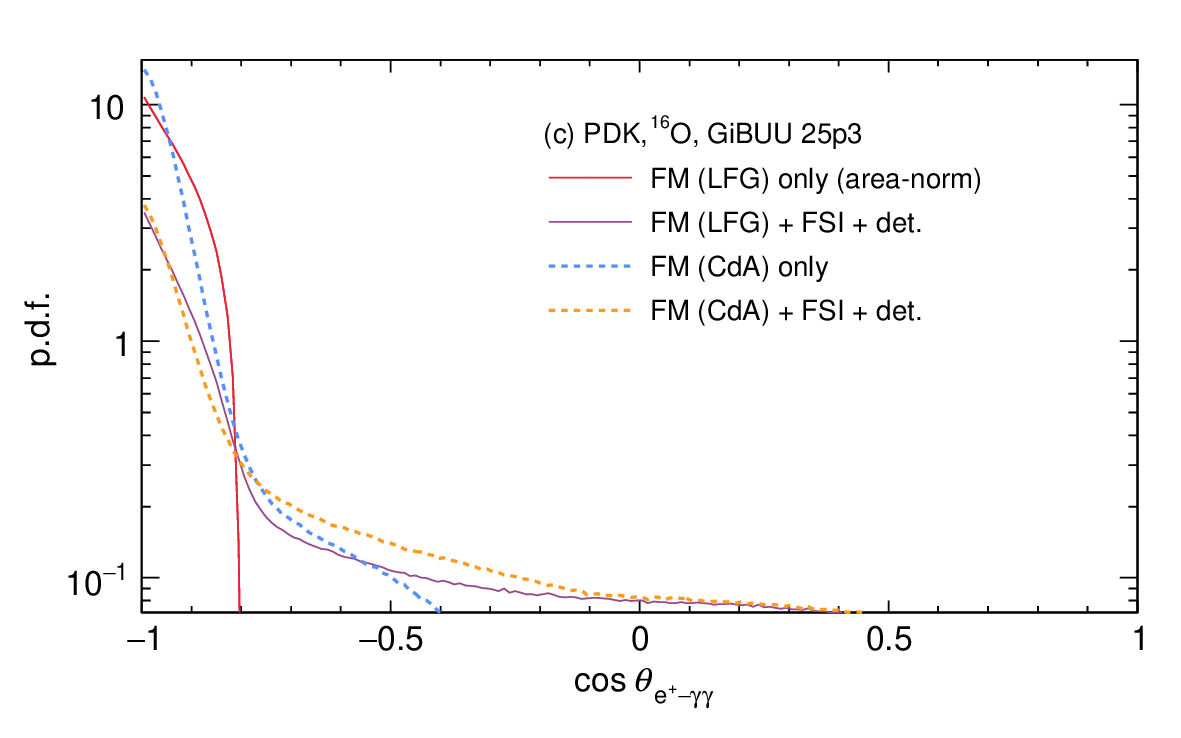}
    \caption{Kinematic distributions of bound proton decay candidate:
        (a) $\megg$, %
        (b) $\pegg$, and %
        (c) opening angle between $\positron$ and $\pi^0$, $\thetaegg$.
    }
    \label{fig:kinematics}
\end{figure}

The invariant mass and total momentum distributions of the $\positron\pi^0$ system for bound proton decay events are shown in \figref{fig:kinematics}a and \figref{fig:kinematics}b, respectively.
When the outgoing $\pi^{0}$ undergoes FSI rescattering and detector smearing, the invariant mass of the system shifts to lower values, producing
a broad shoulder out to large momenta, visible in %
\figref{fig:kinematics}b.
The significant impact of the FSI rescattering and detector smearing on the opening angle between the outgoing positron and $\pi^{0}$ is illustrated in \figref{fig:kinematics}c.
A smaller opening angle is caused by a deviation from a back-to-back topology.

\subsection{Signal detection efficiency and expected background rate}\label{sec:signal_definition}
\begin{figure}[htbp]
    \centering
    \includegraphics[width=\figwid]{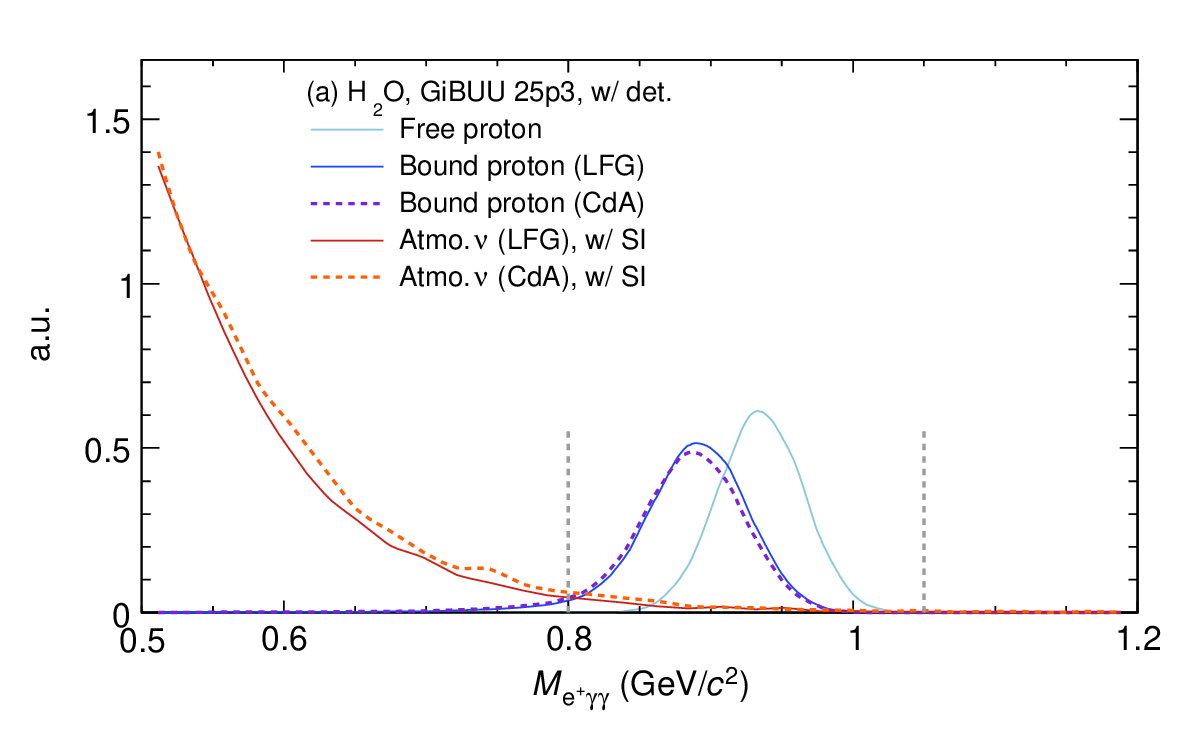}
    \hfill
    \includegraphics[width=\figwid]{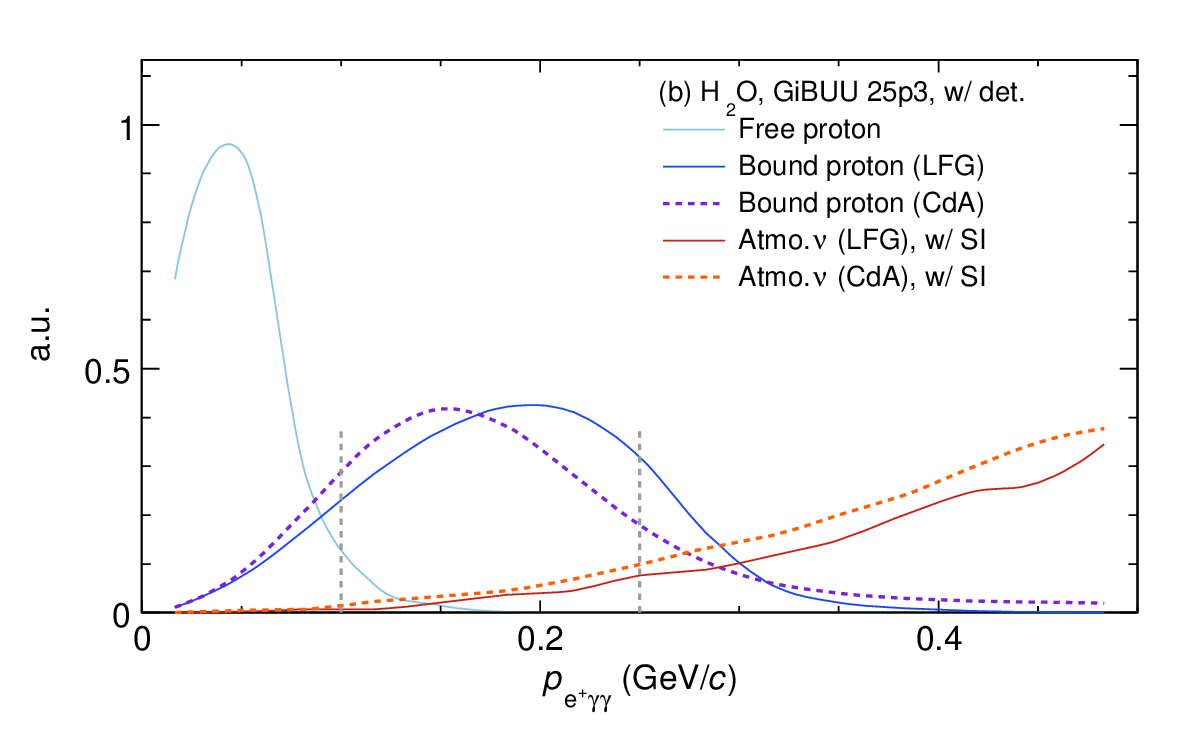}
    \caption{Shape comparison for (a) $M_\egg$ with all selection criteria applied except the $M_\egg$ cut, \textbf{C5},  and (b) $p_\egg$ with all selection criteria applied except the $p_\egg$ cut, \textbf{C6}.
        The signal and background regions are indicated by the vertical dashed lines (see text).
    }
    \label{fig:mass_momentum_shape}
\end{figure}

The following same signal selection criteria as the Super-Kamiokande experiment~\cite{Super-Kamiokande:2020wjk} were applied to the generated proton decay signal and atmospheric neutrino background events to evaluate the signal selection efficiency and residual background rate:
\begin{itemize}
    \item[\textbf{C1}] There must be 2 or 3 identified particles.
    \item[\textbf{C2}] All particles must be showering (e-like).
    \item[\textbf{C3}] There must be no Michel electrons.
    \item[\textbf{C4}] For events with 3 particles, the reconstructed invariant mass of the $\pi^0$ (from two photons) must be within \SI{85}{MeV/\textit{c}^2} to \SI{185}{MeV/\textit{c}^2} (Fig.~\ref{fig:pi0M}).
    \item[\textbf{C5}] The reconstructed invariant mass of the $\positron\pi^0$ system must be within \SI{800}{MeV/\textit{c}^2} to \SI{1050}{MeV/\textit{c}^2} (Figs.~\ref{fig:all} and~\ref{fig:mass_momentum_shape}a).
    \item[\textbf{C6}] The reconstructed total momentum of the $\positron\pi^0$ system must be less than \SI{100}{MeV/\textit{c}}---\textit{Lower} signal region---or  between \SI{100}{MeV/\textit{c}} and \SI{250}{MeV/\textit{c}}---\textit{Upper} signal region (Figs.~\ref{fig:all} and~\ref{fig:mass_momentum_shape}b).
    \item[\textbf{C7}] There must be no tagged neutrons.
\end{itemize}
As shown in Figs.~\ref{fig:all} and~\ref{fig:mass_momentum_shape}b, the reconstructed total momentum of the free proton decay events is mostly below~\SI{100}{MeV/\textit{c}}, which is significantly smaller than that for proton decay events inside oxygen nuclei.
This feature provides strong discrimination against atmospheric neutrino events.
Exploiting this characteristic to further enhance the sensitivity to proton decay searches, two signal regions are defined according to the reconstructed total momentum, following the strategy adopted in previous Super-Kamiokande analyses~\cite{Super-Kamiokande:2016exg}.
The lower signal region is defined by a total momentum below \SI{100}{MeV/\textit{c}}, while
the upper signal region corresponds to a total momentum between \SI{100}{MeV/\textit{c}} and \SI{250}{MeV/\textit{c}}.

For events with three identified particles, %
the $\pi^0$ candidate was identified by selecting the pair of particles with an invariant mass closest to the nominal $\pi^0$ mass of \SI{135}{MeV/\textit{c}^2}. The remaining particle was then assigned as the $\positron$ candidate.

For atmospheric neutrino background events, neutron production often occurs through neutrino interactions or secondary hadronic interactions in water.
The detection of these neutrons, therefore, provides an effective means of background suppression.
Neutrons produced in water are thermalized and subsequently captured by protons, emitting a 2.2~MeV photon.
In the pure-water phase of the Super-Kamiokande detector, as well as in Hyper-Kamiokande, neutrons are detected by observing the faint Cherenkov light induced by this 2.2 MeV photon.
Since the neutron detection efficiency in the Hyper-Kamiokande detector with a photocathode coverage of 20\% was estimated to be approximately 50\%~\cite{Izumiyama:2021pjr}, this analysis assumes a neutron detection efficiency of 50\%$(=\epsilon_{\rm ntag})$, and each event is weighted by the following factor $(w_{\rm ntag})$ according to the true number of captured neutrons $(n_{\rm ncapture})$:
\begin{align}
    w_{\rm ntag} & =(1-\epsilon_{\rm ntag})^{n_{\rm ncapture}}.
\end{align}
This weighting is equivalent to removing events with (at least one) tagged neutrons as done in the latest proton decay search~\cite{Super-Kamiokande:2020wjk}, accounting for the selection criterion \textbf{C7}.

The central values of the signal detection efficiency and expected background rate were obtained using
the generated samples described in Secs.~\ref{sec:eventgen_pdk} and~\ref{sec:eventgen_atm}, with
the default configuration of \gibuu (LFG) without explicit treatment of SRC or in‑medium $\Delta$‑width modifications---the latter are reserved for systematic uncertainty estimation.

As shown in Table~\ref{tab:signalandbkg}, the signal efficiency for proton‑decay events is estimated to be 21.6\% and 17.4\% for lower and upper signal regions, respectively, while the atmospheric neutrino background event rate is estimated to be 0.02~events/(Mton$\cdot$years) and 0.43~events/(Mton$\cdot$years) in the lower and upper signal regions, respectively.
The total signal efficiency is comparable to the value reported by the Super-Kamiokande collaboration (approximately 38\%).
The total background rate without applying the background rejection using the tagged neutron information is estimated to be 1.60~events/(Mton$\cdot$years), and it is consistent with the prediction in the Super-Kamiokande detector, 1.83$\pm0.59$(sys)~events/(Mton$\cdot$years)~\cite{Super-Kamiokande:2020wjk}, as well as the previous estimation using the K2K 1~kton water Cherenkov detector of $1.63^{+0.43}_{-0.33}$(stat)$^{+0.45}_{-0.51}$(sys)~events/(Mton$\cdot$years)~\cite{K2K:2008fnc}.

\begin{table*}[!htpb]
    \centering
    \begin{tabular}{ccccc}
        \toprule
        \gibuu  variation                & \makecell{Signal efficiency                      \\--Lower (\%)} &\makecell{Signal efficiency\\--Upper (\%)} &\makecell{Bkg. rate\\--Lower} &\makecell{Bkg. rate\\--Upper}  \\
        \midrule  \midrule
        \makecell{default}               & 21.6                        & 17.4 & $0.022\pm 0.003$ & $0.43\pm 0.02$ \\ \midrule
        \makecell{\cda}                  & 22.1                        & 16.1 & 0.040 & 0.72 \\ \midrule
        \makecell{Oset}                  & 21.7                        & 18.4 & 0.029 & 0.39 \\ \midrule
        \makecell{
            In-med.
            $\alpha=1.2$ }
                                         & 21.6                        & 17.5 & 0.029 & 0.48 \\ \midrule
        \makecell{ In-med. $\alpha = 2$} & 21.6                        & 17.6 & 0.035 & 0.56 \\
        \bottomrule
    \end{tabular}
    \caption{%
        The signal detection efficiencies and expected atmospheric neutrino background rates, expressed in units of Mton$\cdot$years, under different \gibuu variations.
        Here, ``Lower'' and ``Upper'' refers to the lower and higher signal regions, respectively (see \textbf{C6} in Sec.~\ref{sec:signal_definition}).
        The statistical errors for the background event rate with the default \gibuu configuration are given; those for the other configurations are expected to be of similar magnitude. The statistical errors of the signal efficiency are less than 0.001\%, and therefore are not shown. All statistical errors correspond to the  arbitrarily generated MC statistics used. 
    }
    \label{tab:signalandbkg}
\end{table*}

\subsection{Systematic uncertainties}\label{sec:systematics}

The uncertainties associated with nuclear effects were evaluated by varying the nuclear model configurations in \gibuu (see Section~\ref{sec:gibuu}).
Specifically, we consider alternative configurations that incorporate different treatments of the initial-state nuclear model and FSIs.

For the initial-state nuclear model, we compare LFG with the \cda model.
The differences between the two models arise primarily from the distributions of the nucleon binding energy and the initial nucleon momentum.
The high momentum tail from \cda significantly affects the kinematic distributions of the decay products, particularly the momentum distribution and the opening angle distribution, and therefore the invariant mass, of the $\positron$ and $\pi^0$ system, as illustrated in \figref{fig:kinematics}.

The impact of these differences for the proton decay signal events is reflected in the reconstructed momentum of the $\pi^{0}$ and of the total momentum of the system, as can be seen in~\figref{fig:pi0p} and~\figref{fig:mass_momentum_shape}.
The softer spectrum by the CdA model has a larger fraction of events below the signal cut, resulting in a higher efficiency.

\begin{figure}[htbp]
    \includegraphics[width=\figwid]{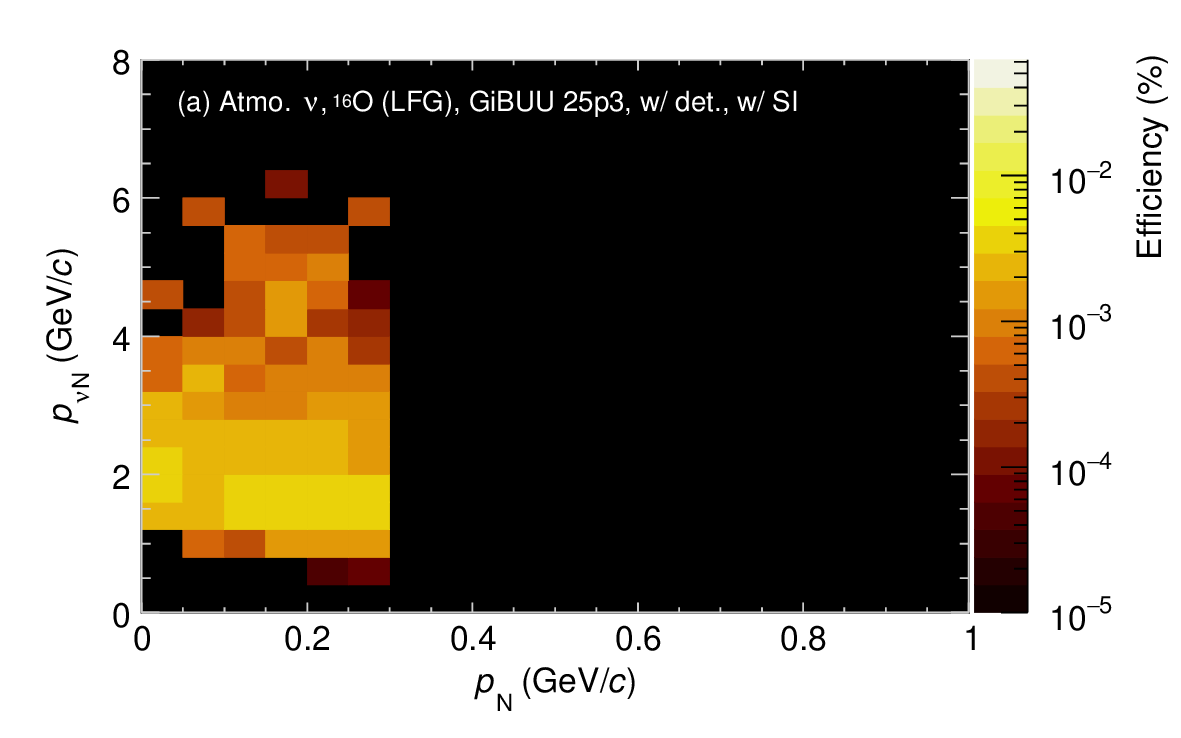}
    \includegraphics[width=\figwid]{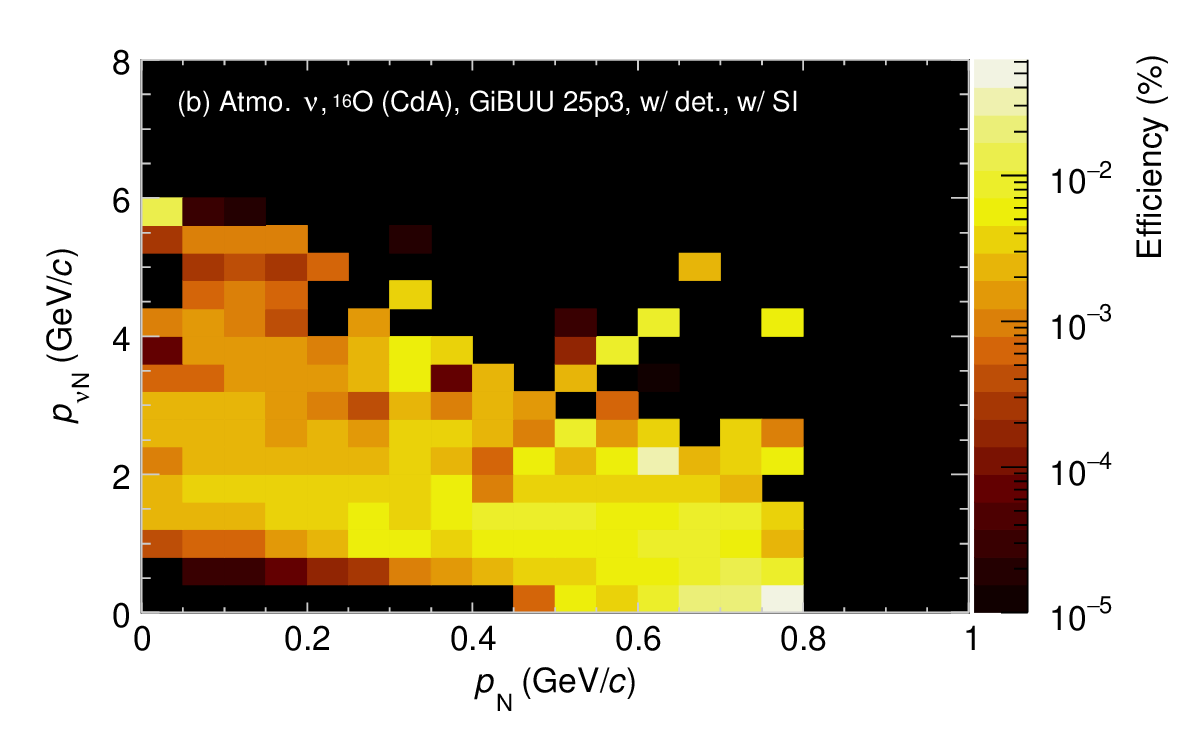}
    \caption{
        Selection efficiency for atmospheric neutrino background events
        as a function of the true initial nucleon momentum ($p_\nucleon$) and the true total momentum of the initial neutrino–nucleon system ($p_{\nu\nucleon}$), for (a) LFG and (b) \cda.
        The efficiency is defined as the fraction of generated events that pass the selection criteria up to \textbf{C6}.
    }
    \label{fig:eff_bkg_2d}
\end{figure}

For the atmospheric neutrino background events, the expected number of background events is significantly enhanced in the \cda model (Fig.~\ref{fig:mass_momentum_shape}) due to the high-momentum tail of the initial nucleon momentum distribution.
As is shown in Fig.~\ref{fig:eff_bkg_2d}, the high-momentum tail can effectively reduce the initial total momentum of the neutrino–nucleon system.
As a result, in proton decay searches that select events with low reconstructed total momentum, such events are more likely to be misidentified as signal candidates.

It should be noted that the latest proton decay search by Super-Kamiokande employed the relativistic Fermi gas model~\cite{Smith:1972xh}, which does not include a high-momentum tail in the distribution.
Variations in the atmospheric neutrino background associated with differences in the nucleon momentum distribution were not evaluated, and no corresponding systematic uncertainty was assigned in that analysis.
The present study indicates that such effects could influence the estimated background rate and should be considered in future proton decay searches.

\begin{figure}[!htbp]
    \centering
    \includegraphics[width=\figwid]{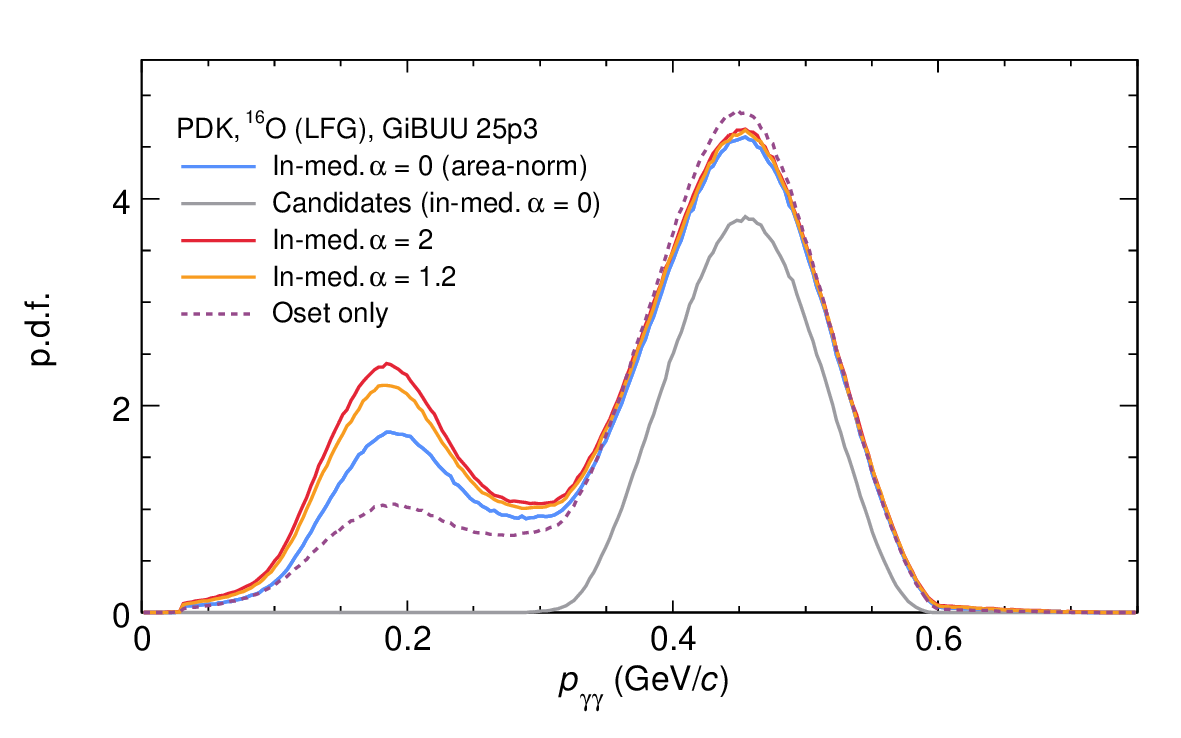}
    \caption{
        $\pgg$ distribution from proton decay events after FSI but before detector smearing, compared to that from the post-cut candidates where detector effects have been included.
        The in-medium modifications controlled by the $\alpha$ parameter (Eq.~\ref{eq:Delta-supp} where $\alpha=0$ means no modification which is the default choice) are compared to Oset broadening alone.
    }\label{fig:pi0_inmed}
\end{figure}

To illustrate the impact from the pion FSIs, Figure~\ref{fig:pi0_inmed} compares the $\pgg$ distribution from proton decay events under different FSI configurations.
The main peak at around $\SI{450}{MeV/\textit{c}}$ corresponds to events that experience no or only minor FSI and constitutes the dominant contribution to the selected signal candidates.
The impact of the in-medium modifications
is estimated by varying the $\alpha$ parameter in Eq.~(\ref{eq:Delta-supp}), among 0, 1.2, and 2 within the \gibuu simulation.
Increasing $\alpha$ reduces the $\pi^0$ absorption probability, leading to a more pronounced secondary peak around $\SI{200}{MeV/\textit{c}}$, which arises from energy dissipation through FSIs relative to the free-proton decay peak of \decayp.
Oset broadening increases the signal detection efficiency by reducing the cross section for $\pi$ FSIs mediated through the $\Delta$ resonance.
Different strengths are observed with the Oset broadening and in-medium modifications but their impact is not significant for the selected candidates.

\begin{figure}[!htbp]
    \centering
    \includegraphics[width=\figwid]{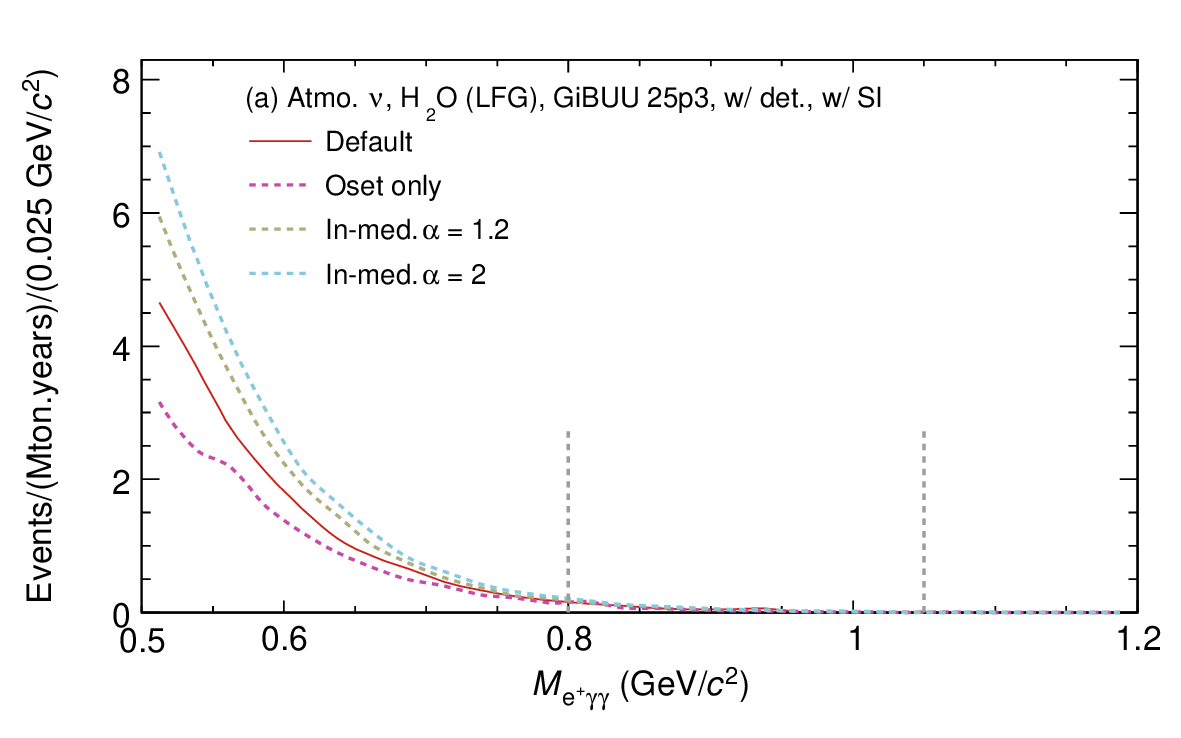}
    \includegraphics[width=\figwid]{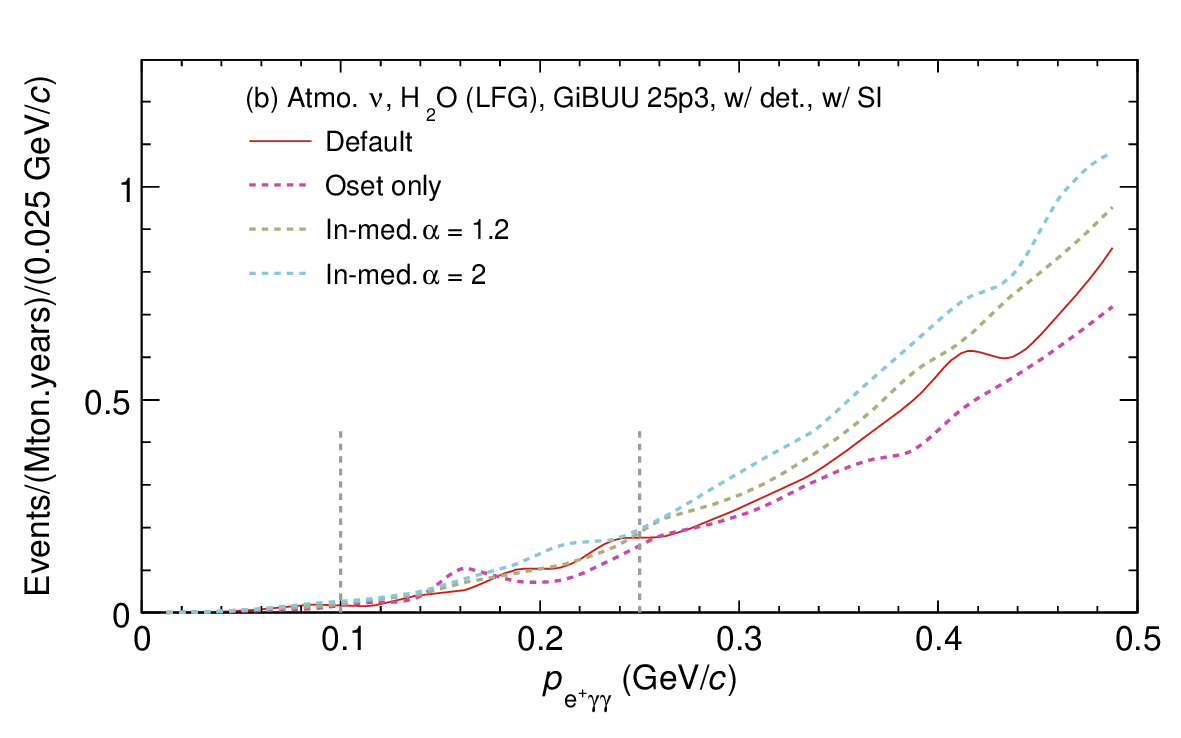}
    \caption{
        $\megg$ and $\pegg$ distributions
        comparing different pion FSIs. The signal and background regions are indicated by the vertical dashed line.
    }\label{fig:background_inmed}
\end{figure}

In addition, as shown in Fig.~\ref{fig:background_inmed} for atmospheric neutrino interaction background,
a larger value of $\alpha$ reduces the pion absorption probability in FSIs and consequently increases the expected background rate, while
Oset broadening slightly decreases it by suppressing the cross section of neutrino interactions accompanied by $\pi$ production, which constitutes the dominant background channel for the $\pdk$ channel.

The corresponding changes in the signal detection efficiency and the background event rate are summarized in Tables~\ref{tab:signalandbkg}.
Regardless of whether the pion undergoes scattering or absorption, a $\pi^{0}$ produced in proton decay rarely remains within the signal region after experiencing FSIs.
Consequently, the variation in the signal detection efficiency due to in-medium effects is small.

\begin{table*}[!tb]
    \centering
    \begin{tabular}{c@{\hspace{0.5cm}}c@{\hspace{0.5cm}}c@{\hspace{0.5cm}}c@{\hspace{0.5cm}}c@{\hspace{0.5cm}}c} \hline
        Item                               & \makecell{FM and SRC} & \makecell{$\pi$ FSI and                             \\ $\nu$ interaction} & $\nu$ flux & Reconstruction & Total \\   \midrule  \midrule
        Signal detection efficiency--Lower & 2.3 (7.4)             & 0.3 (2.9)               & N.A. & 3.6  & 4.3 (8.8)   \\
        Signal detection efficiency--Upper & 7.5 (10.6)            & 2.7 (11.7)              & N.A. & 3.2  & 8.6 (16.1)  \\
        Atmospheric neutrino backgrounds   & 65.8 (N.A.)           & 16.1 (23.5)             & 7.3  & 20.5 & 71.1 (32.0) \\ \hline
    \end{tabular}
    \caption{
        Summary of the systematic uncertainties (\%) on the signal detection efficiency and the expected atmospheric neutrino background rate.
        Here, ``Lower'' and ``Upper'' refers to the lower and higher signal regions, respectively (see \textbf{C6} in Sec.~\ref{sec:signal_definition}).
        In the same manner as Ref.~\cite{Super-Kamiokande:2020wjk}, the systematic uncertainty on the expected atmospheric neutrino background rate is common for both signal regions.
        In this \gibuu-based analysis, the systematic uncertainty associated with FM and SRCs was evaluated by comparing the LFG and \cda models.
        The systematic uncertainties related to $\pi$ FSIs and neutrino interactions were estimated by introducing Oset broadening and in-medium modifications.
        For comparison, the corresponding systematic uncertainties from the latest Super-Kamiokande proton decay search~\cite{Super-Kamiokande:2020wjk} are shown in parentheses.
        For the signal detection efficiency, the uncertainties assigned to ``Correlated decay'' and ``Fermi momentum'' in Table~VI of Ref.~\cite{Super-Kamiokande:2020wjk} are combined in quadrature and treated as the systematic uncertainty associated with the Fermi momentum and short-range correlations.
        Similarly, for the background event rate, the uncertainties associated with ``$\pi$ FSI'' and ``Neutrino interaction'' in Table~VII of Ref.~\cite{Super-Kamiokande:2020wjk} are combined in quadrature.
        As noted in the main text, the Super-Kamiokande analysis did not evaluate the atmospheric neutrino background uncertainty arising from differences in the Fermi momentum distribution and short-range correlations.
        The uncertainties related to event reconstruction and the neutrino flux are taken from Tables~VI and~VII of Ref.~\cite{Super-Kamiokande:2020wjk}.
    }
    \label{tab:summary_systematics}
\end{table*}

In this analysis, the treatment of systematic uncertainties is designed to closely follow that adopted in the Super-Kamiokande proton decay searches~\cite{Super-Kamiokande:2020wjk}.
The systematic uncertainties associated with nuclear effects are separated into two categories: those related to FM and SRCs, and those related to $\pi$ FSIs and neutrino interactions.
The uncertainty arising from the FM and SRCs was evaluated from the relative difference in the signal detection efficiency and atmospheric neutrino background event rate obtained with the LFG and \cda models.
The uncertainty associated with $\pi$ FSIs and neutrino interactions was estimated by taking the unbiased standard deviation of the signal detection efficiency and background event rate obtained from the three different implementations of Oset broadening and in-medium modifications listed in Table~\ref{tab:signalandbkg}.
Other systematic uncertainties, such as those related to the event reconstruction and the neutrino flux, are taken to be the same as those used in the latest Super-Kamiokande analysis~\cite{Super-Kamiokande:2020wjk}.
Although a direct comparison is not strictly possible because of the differences in the models and procedures employed for the evaluation of systematic uncertainties between the two analyses, Table~\ref{tab:summary_systematics} summarizes the systematic uncertainties adopted in the present proton decay sensitivity estimate discussed in Sec.~\ref{sec:sensitivity_projections} using the \gibuu model, along with those reported in the Super-Kamiokande analysis for reference.

\subsection{Search sensitivity projections}\label{sec:sensitivity_projections}
The sensitivity is calculated within a Bayesian framework. For a hypothetical proton decay width $\Gamma$ into the $\positron\pi^0$ channel, the probability of observing $n$ events in a given exposure is evaluated, where $n$ is an assumed event count varied in the sensitivity study. This probability is described by the following Poisson likelihood:
\begin{equation}
    \poissp(n_{i}|\Gamma;\lambda_{i},\epsilon_{i},b_{i}) = \frac{e^{-(\Gamma \lambda_{i} \epsilon_{i} + b_{i})}(\Gamma \lambda_{i} \epsilon_{i} + b_{i})^{n_{i}}}{n_{i}!}, \label{eq:poisson}
\end{equation}
where $i=1,~2$ denotes the ``Lower'' and ``Upper'' signal region, respectively, $\lambda$ is the detector exposure in proton-years, $\epsilon$ is the signal detection efficiency, and $b$ is the expected number of background events. The uncertainties in $\lambda$, $\epsilon$, and $b$ are incorporated through prior distributions $\poissp(\lambda)$, $\poissp(\epsilon)$, and $\poissp(b)$, respectively. Using Bayes' theorem, the posterior probability distribution for $\Gamma$ given $n_{i}$ observed events is
\begin{align}
    \poissp(\Gamma|n_{i}) = & \frac{1}{A_{i}}\iiint \dd\lambda_{i} \dd\epsilon_{i} \dd b_{i} \, \poissp(n_{i}|\Gamma;\lambda_{i},\epsilon_{i},b_{i}) \notag \\
                            & \times \poissp(\lambda_{i})\poissp(\epsilon_{i}) \poissp(b_{i}) \poissp(\Gamma) , \label{eq:posterior}
\end{align}
with a normalized flat prior $\poissp(\Gamma)$ for $\Gamma \geq 0$, and a normalization constant $A_i$.

The prior $\poissp(\lambda)$ is conservatively modeled as a Gaussian distribution centered on the nominal exposure, with a systematic uncertainty of 1\%.
The prior $\poissp(\epsilon)$ is taken as a Gaussian distribution centered on the signal efficiency in the \gibuu configuration shown in Table~\ref{tab:signalandbkg}, with a width given by its systematic uncertainty presented in Table~\ref{tab:summary_systematics} and truncated at zero.
The raw number of remaining atmospheric neutrino background events in the simulation is limited.
Therefore, to account for both the statistical fluctuations of the remaining background and the associated systematic uncertainties, the prior for the background rate $\poissp(b)$ is modeled as a convolution of a Poisson distribution and a Gaussian distribution~\cite{Super-Kamiokande:2009yit, Super-Kamiokande:2016exg, Super-Kamiokande:2020wjk}.
The Gaussian is centered at the expected background yield and has a width corresponding to the systematic uncertainty.
The 90\% credible interval upper bound
on the proton decay width, $\glim\left(n_\los,~n_\ups\right)$, is determined by solving
\begin{equation}
    \int_0^{\glim}\prod_{i} \poissp(\Gamma|n_{i}) \dd\Gamma = 90\%. %
\end{equation}

The sensitivity projection assumes that the number of observed events
is all contributed by the background, representing the expected sensitivity in the absence of a true signal.
The expected  upper
limit on the proton decay width is obtained by averaging over the Poisson distribution of possible background counts:
\begin{align}
    \gexp & = \sum_{n_\los=0}^\infty\sum_{n_\ups=0}^\infty \frac{e^{-b_\los} b_\los^{n_\los}}{n_\los!} \frac{e^{-b_\ups} b_\ups^{n_\ups}}{n_\ups!}
    \glim(n_\los, n_\ups). \label{eq:gamma_expect}
\end{align}
The corresponding expected lower limit on the proton lifetime is then $\texp = 1/\gexp$.

\begin{figure}[htbp]
    \centering
    \includegraphics[width=\figwid]{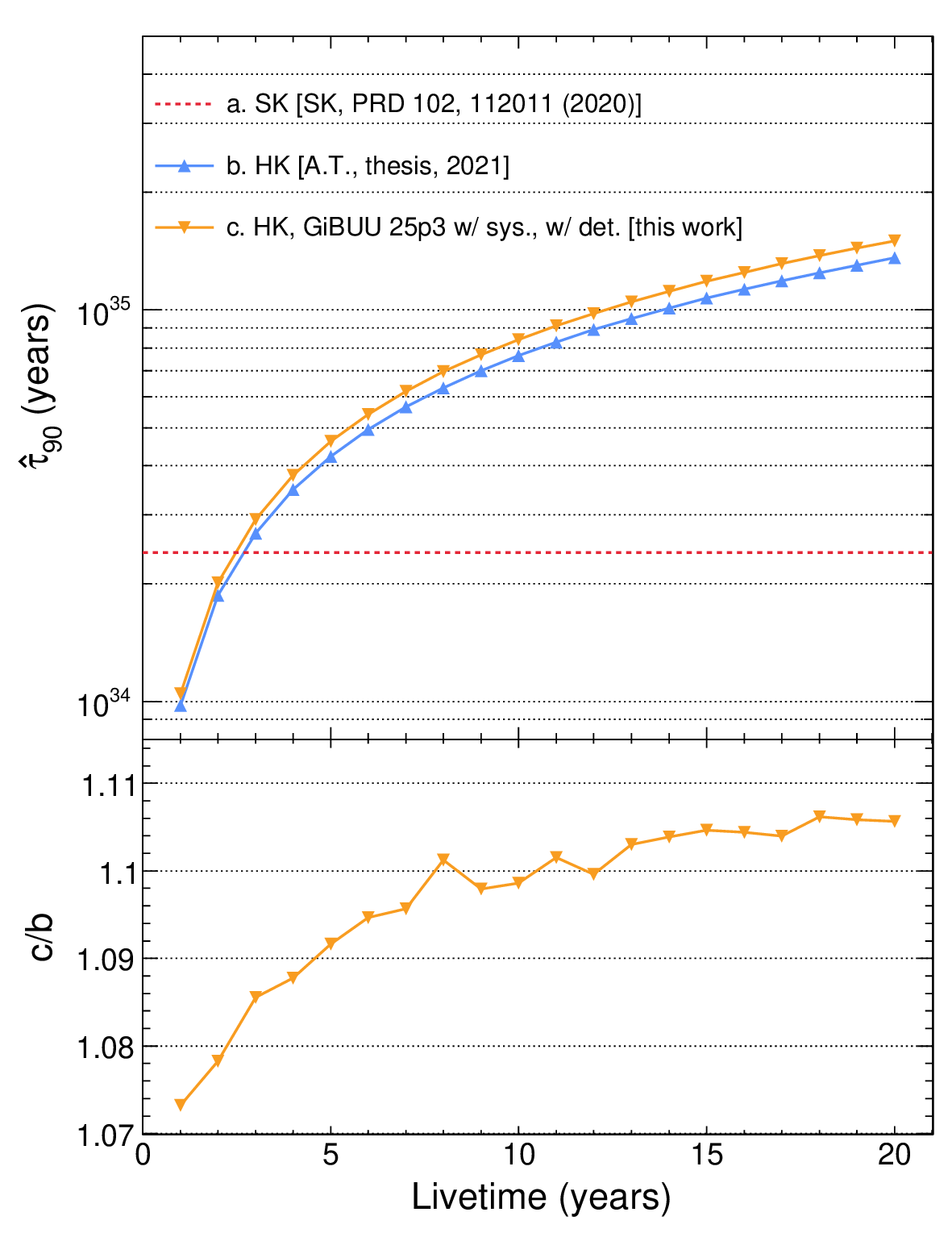}
    \caption{The projected 90\% credible interval lower limit, $\texp$, on the proton lifetime for the $\pdk$ channel as a function of exposure time for the Hyper-Kamiokande experiment.
        The red horizontal dashed line (a) represents the current lifetime limit obtained in Ref.~\cite{Super-Kamiokande:2020wjk}, $2.4\times10^{34}$~years.
        The blue upward triangles (b) represent the expected search sensitivity in Hyper-Kamiokande, evaluated in Ref.~\cite{Takenaka:2020cmb}.
        The orange downward triangles (c) denote the search sensitivity in Hyper-Kamiokande with the derived search performance using the \gibuu simulation in this work.
        The ratio between the curves (c) and (b) is plotted in the lower panel.
    }\label{fig:proton_lifetime_limit}
\end{figure}

The projected 90\% credible sensitivity on the proton lifetime for the $\pdk$ channel as a function of exposure time for the Hyper-Kamiokande experiment is shown in \figref{fig:proton_lifetime_limit}.
Although differences in the adopted nuclear interaction models lead to variations in the systematic uncertainties associated with the signal detection efficiency and the expected background event rate, as shown in Table~\ref{tab:summary_systematics}, the central values of the signal efficiency and background rate themselves closely reproduce the performance reported in the Super-Kamiokande analysis~\cite{Super-Kamiokande:2020wjk}.
Consequently, the resulting proton decay search sensitivity is largely comparable to previous estimates in Refs.~\cite{Hyper-Kamiokande:2018ofw, Takenaka:2020cmb}.
The observed 7 to 10\% difference in sensitivity is primarily attributed to the slightly higher signal detection efficiency and lower systematic uncertainty on it in the present analysis using the \gibuu model.

\section{Summary and outlook}\label{sec:conclusion}
In this work, we have reassessed the impact of nuclear effects on the sensitivity of next-generation proton decay searches in the $\pdk$ channel. Using the \gibuu model, which provides a consistent treatment of nuclear effects for both the proton decay signal and the atmospheric neutrino background, we have evaluated the signal detection efficiency, background event rate, and systematic uncertainties associated with nuclear effects.
By implementing a particle kinematic reconstruction performance model that emulates a water Cherenkov detector, we find that the use of the \gibuu model reproduces signal detection efficiencies and expected background event rates comparable to those obtained in previous proton decay searches in the Super-Kamiokande experiment~\cite{Super-Kamiokande:2020wjk}.
As discussed below, although the systematic uncertainties estimated in this analysis exhibit some differences from those reported by Super-Kamiokande, the resulting proton decay search sensitivity under Hyper-Kamiokande conditions is largely consistent with previous estimates in Refs.~\cite{Hyper-Kamiokande:2018ofw, Takenaka:2020cmb} at a level of around 10\%.
Ultimately, \gibuu provides the self-consistent modeling required to move beyond \textit{ad hoc} simulations, offering a valuable framework for characterizing the complex nuclear landscapes of the $10^{35}$-year lifetime regime.

In this analysis, two models describing the Fermi momentum distribution, the local Fermi gas (LFG) model and the Cioffi degli Atti and Simula (\cda) model, are compared to evaluate their impact on the detection efficiency of proton decay events in oxygen nuclei and on the background event rate arising from neutrino interactions on bound nucleons in oxygen.
The introduction of the high-momentum component associated with short-range correlations described by the \cda model leads to a moderate change in the signal detection efficiency at the level of about 8\%, while it induces a significant increase in the atmospheric neutrino background rate of nearly 70\%.
It would therefore be desirable to take this effect into consideration in future proton decay searches.
In the context of the systematics connected to FSIs, complicated mechanisms dominate the pion rescattering inside the nucleus and possible medium modifications to the intermediate resonances.
We have shown that these cause visible distortions in the reconstructed mass and momentum distributions, and these effects lead to variations of a few percent in the signal detection efficiency and to an uncertainty of approximately 16\% in the expected atmospheric neutrino background rate.
The same process already showed a significant impact on neutrino-nucleus interactions~\cite{Mosel:2023zek,Yan:2025aau}.

Addressing these sources of uncertainty requires a multi-probe approach to establish a connection between different processes governed by the same underlying nuclear dynamics: electron-nucleus scattering, neutrino-nucleus scattering, and proton decay. This interconnectedness ensures that proton decay searches are constrained by high-precision benchmarks from the broader scattering community.
A critical advantage of this framework is the unified modeling of medium effects across both signal and background processes.
This consistency will enable, for the first time, the rigorous consideration of correlations between signal and background systematic uncertainties, which is the factor currently neglected in both this study and existing experimental searches.
Accounting for these correlations is essential for a truly unbiased extraction of the proton lifetime.
In addition to the $\pdk$ channel investigated in this work, several other decay modes with relatively large predicted branching ratios arise in various GUT models, such as $\textrm{p}\rightarrow\mu^{+}\pi^{0}$, $\textrm{p}\rightarrow\nu\textrm{K}^{+}$, $\proton \to \mu^+ \rm K^0$.
It would be interesting to extend the present methodology to these alternative channels to assess the impact of nuclear effects and associated systematic uncertainties in a consistent framework.

\begin{acknowledgments}
    The authors appreciate Seungho Han for providing the information regarding the secondary interaction in water Cherenkov detectors and \geant setup, as well as commenting on our manuscript.
    Q.Y. and X.L. thank Wanlei Guo and Zhenning Qu for discussions.
    Q.Y. and Y.Z. are supported by National Natural Science Foundation of China (NSFC) under contract 12221005.
\end{acknowledgments}

\bibliography{bibliography}%

\appendix

\section{Individual ternary classification plots}\label{app:dec}

\begin{figure}[!htbp]
    \centering
    \includegraphics[width=\figwid]{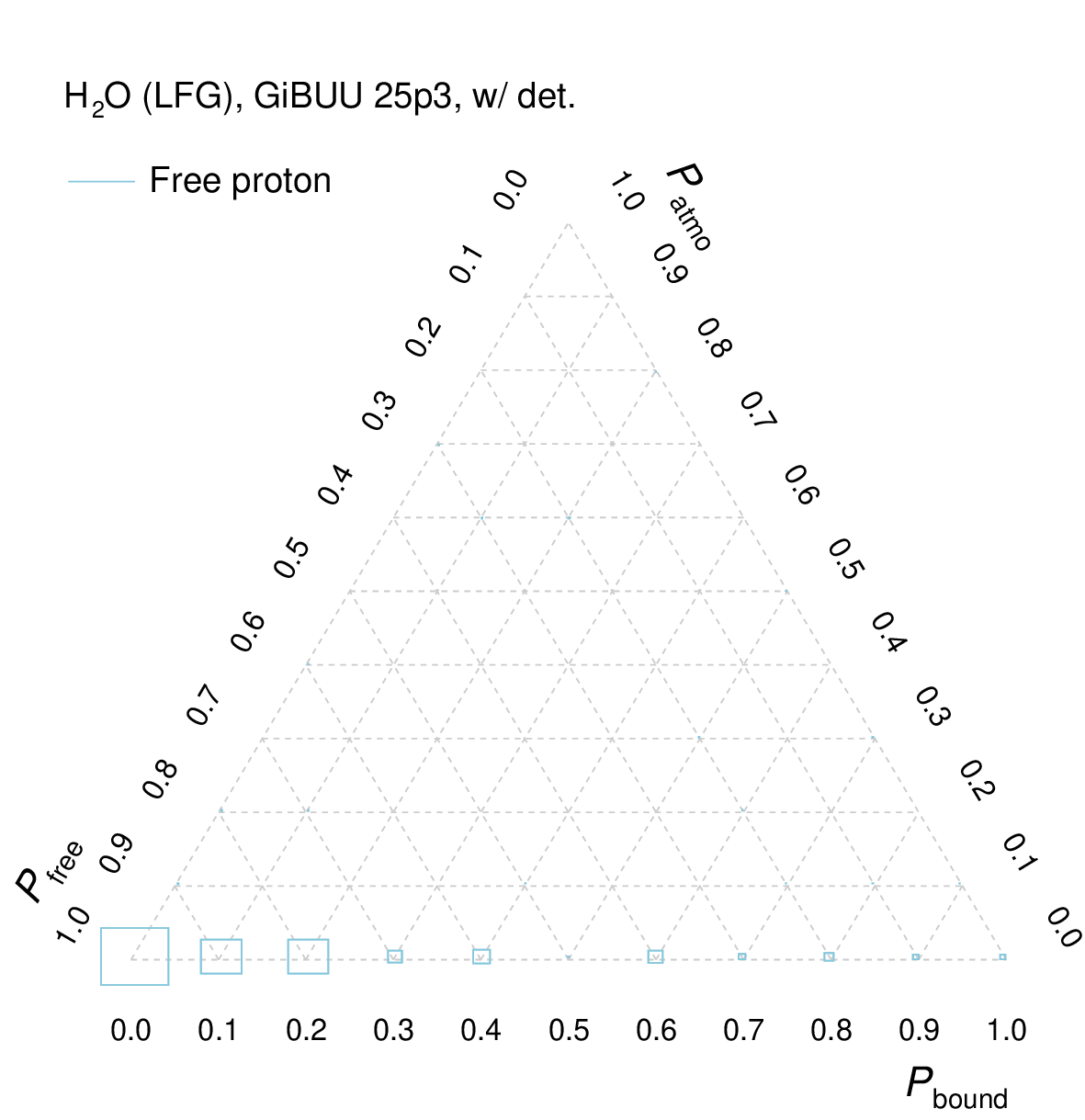}
    \caption{
        Ternary classification of free proton decay events.
        The events are concentrated near the $\pfree$ vertex, indicating they are correctly classified as signal and well separated from the atmospheric neutrino background region.
    }\label{fig:plot_free_proton}
\end{figure}

\begin{figure}[!htbp]
    \centering
    \includegraphics[width=\figwid]{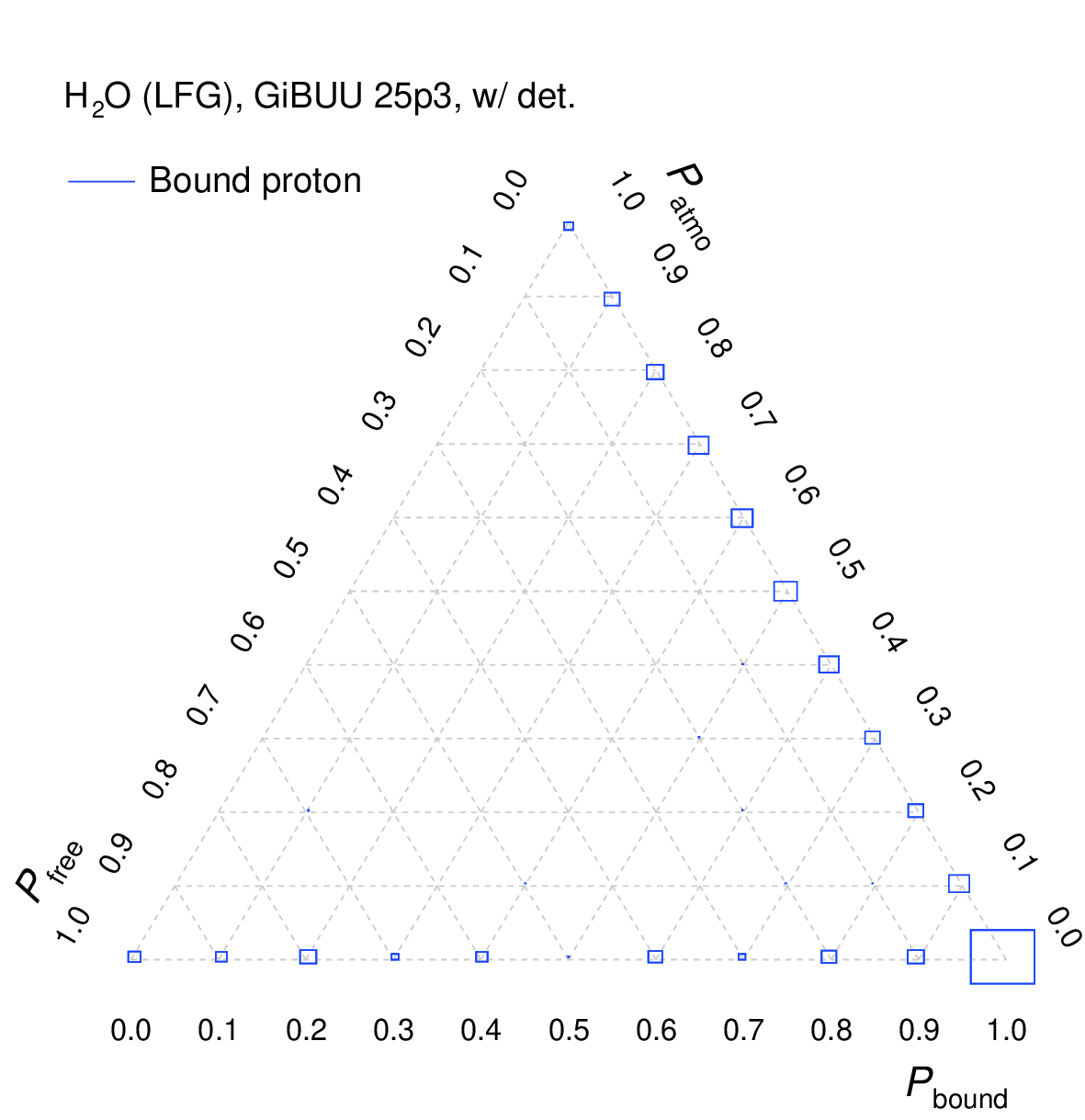}
    \caption{
        Ternary classification of bound proton decay events. Compared to free proton decay, these events show a broader spread toward the $\patmo$ vertex, reflecting the increased similarity to atmospheric neutrino background due to nuclear effects.
    }\label{fig:plot_bound_proton}
\end{figure}

\begin{figure}[!htbp]
    \centering
    \includegraphics[width=\figwid]{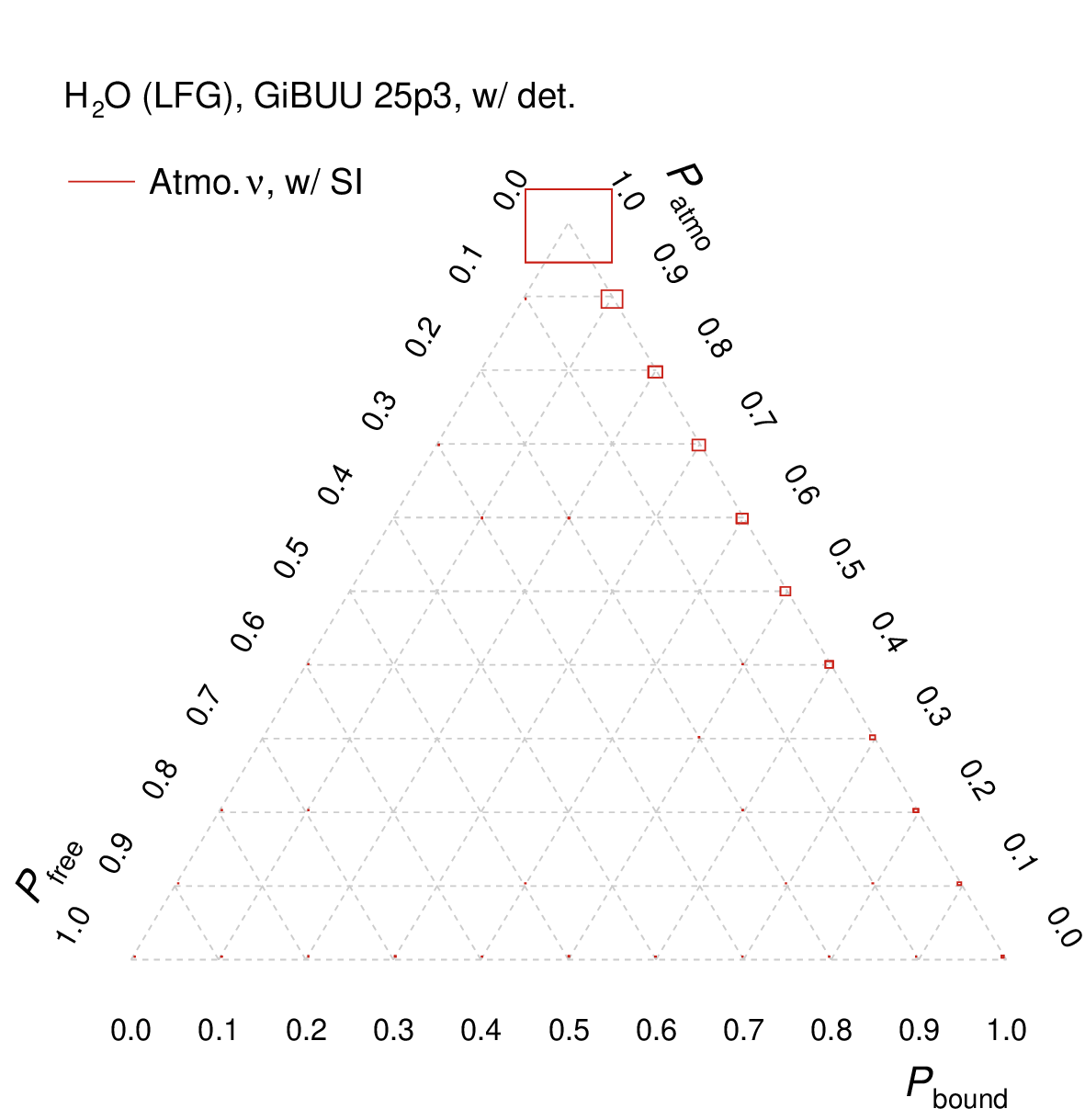}
    \caption{
        Ternary classification of atmospheric neutrino background events.
        A significant fraction of background events populate the region near the $\patmo$ vertex, but a non-negligible tail extends into the bound proton decay signal region, illustrating the source of background contamination for proton decay.
    }\label{fig:plot_background}
\end{figure}

Figures~\ref{fig:plot_free_proton}-~\ref{fig:plot_background} decompose the combined ternary plot shown in \figref{fig:3sample} into its constituent event categories. They visually confirm the hierarchical separation: free proton decay events are isolated near their corresponding vertex with minor mixing with bound proton, while bound proton decay and atmospheric neutrino background events exhibit increasing overlap.

\begin{figure}[!htbp]
    \centering
    \includegraphics[width=\figwid]{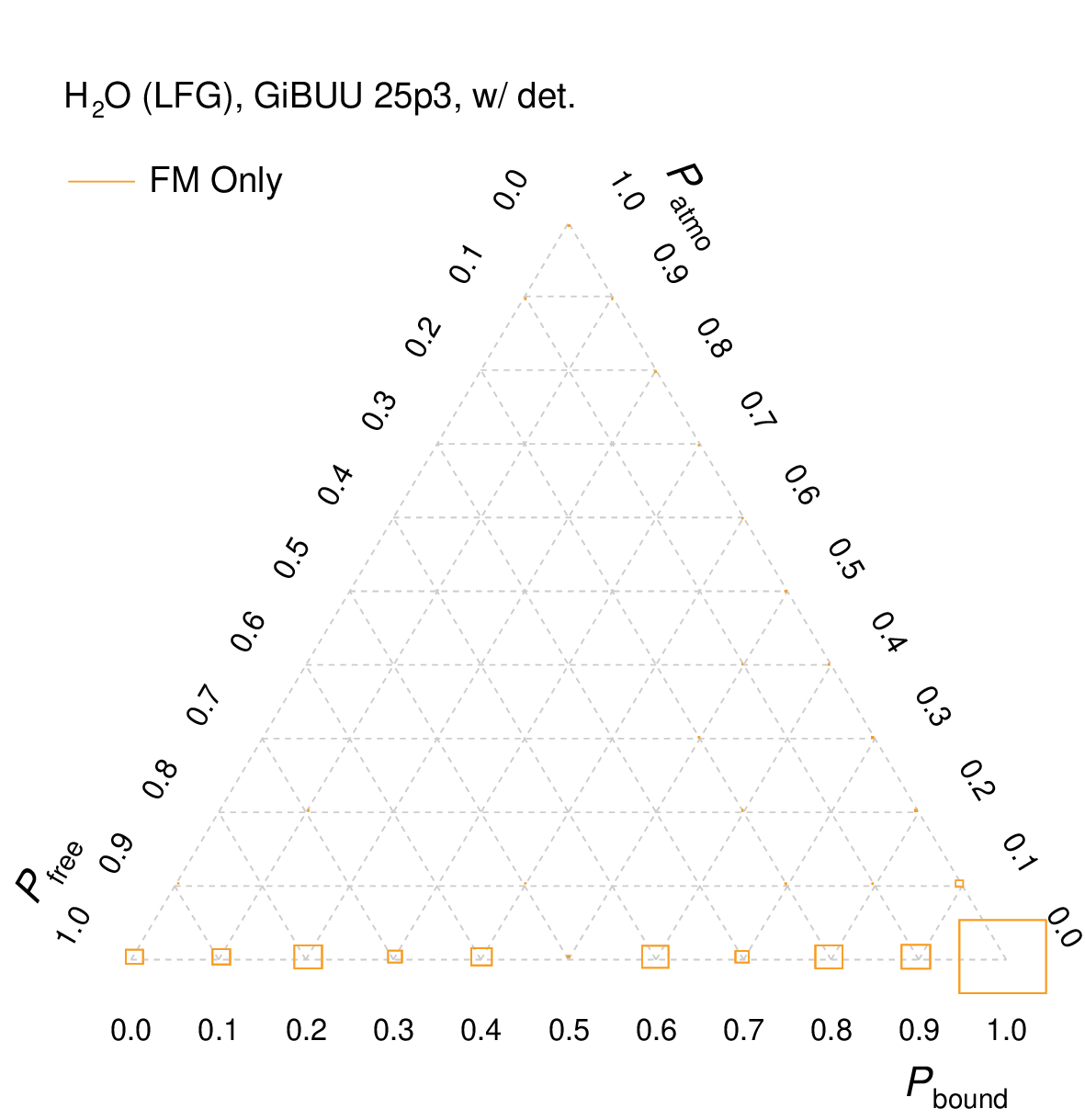}
    \caption{
        Ternary classification of bound proton decay events affected only by FM without SRC or FSIs. These events remain largely confined to the proton decay signal region---partly classified as free-proton decay---showing minimal mixing with the atmospheric background.
    }\label{fig:split_1}
\end{figure}

\begin{figure}[!htbp]
    \centering
    \includegraphics[width=\figwid]{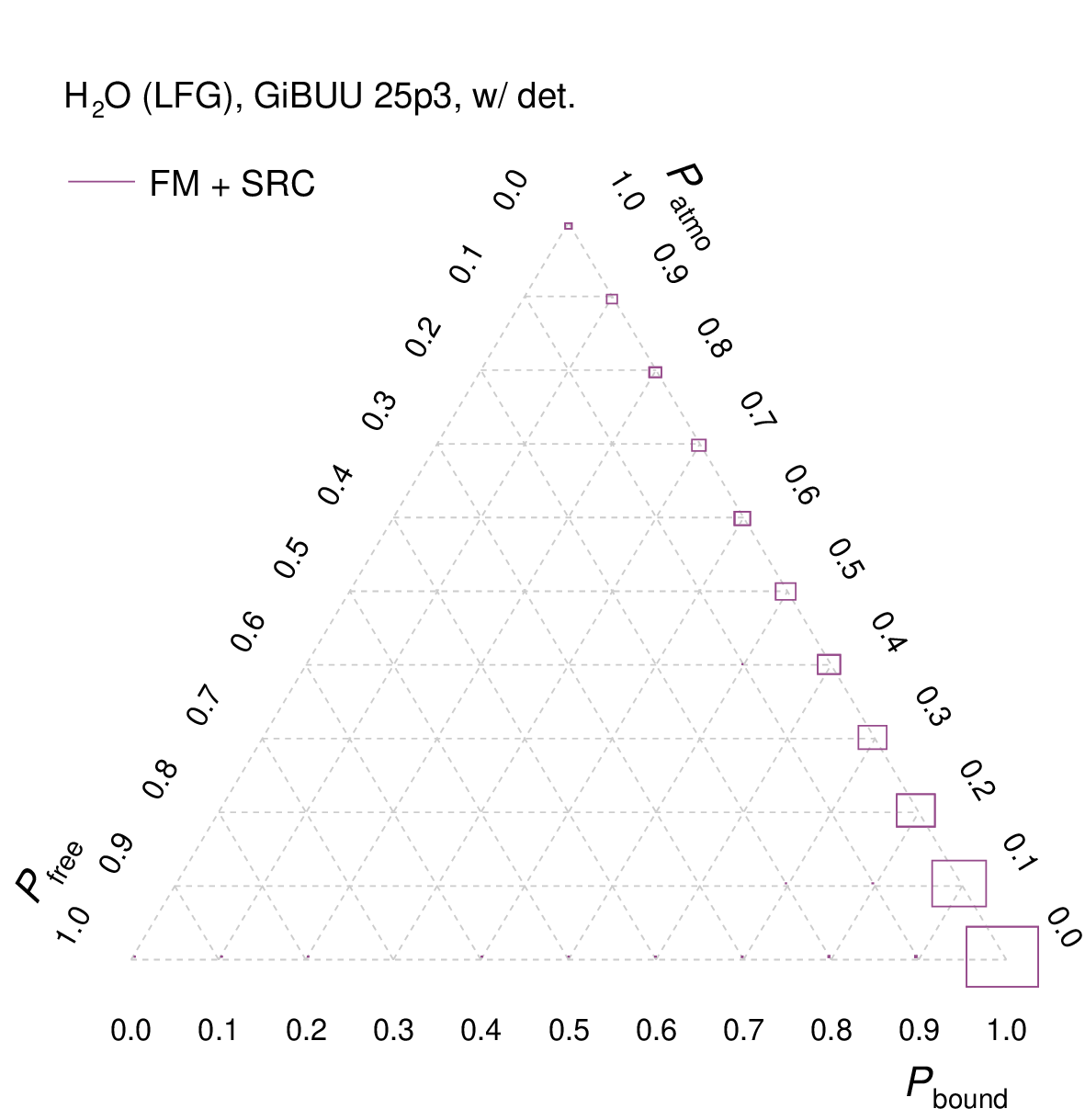}
    \caption{
        Ternary classification of bound proton decay events affected by both FM and SRC. The inclusion of SRC effects begins to shift the event distribution slightly toward the atmospheric background region, though the majority of events are still classified as signal.
    }\label{fig:split_2}
\end{figure}

\begin{figure}[!htbp]
    \centering
    \includegraphics[width=\figwid]{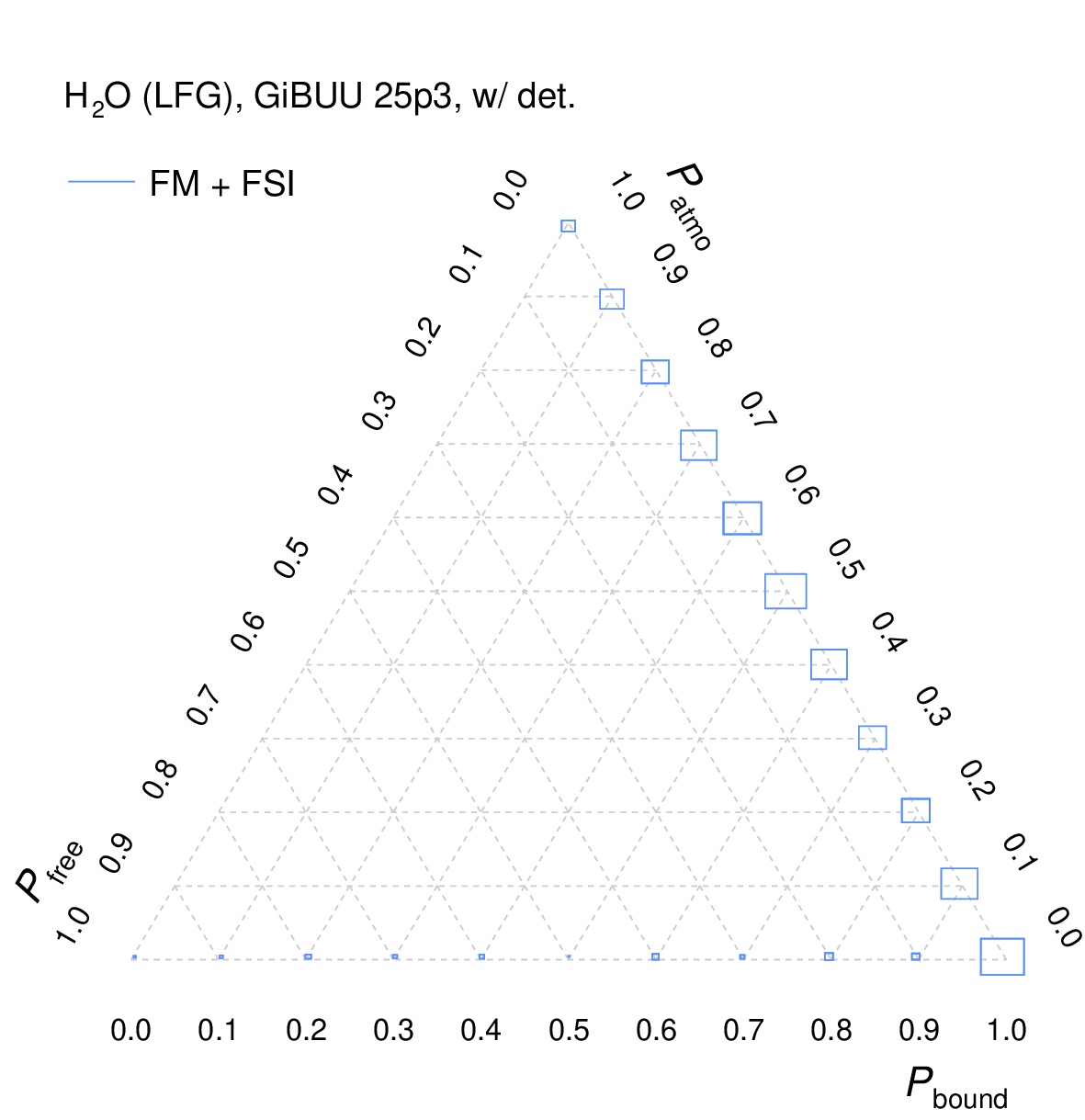}
    \caption{
        Ternary classification of bound proton decay events affected by FM and pion FSIs. The inclusion of pion FSIs causes a significant migration of events into the atmospheric neutrino background region, demonstrating that FSI is the dominant nuclear effect responsible for the reduced signal-background separation.
    }\label{fig:split_3}
\end{figure}

The decomposed plots, Figures~\ref{fig:split_1}-\ref{fig:split_3}, illustrate the incremental impact of different nuclear effects on the classification of bound proton decay events, as also summarized in \figref{fig:split_oxygen}. The progressive overlap with the atmospheric background region underscores the critical role of final-state interactions in shaping the detectable signal topology and the associated systematic uncertainties.

\newpage


\section{\cda initial state}\label{app:initial_state}

\begin{figure}
    \centering
    \begin{subfigure}{\figwid}
        \centering
        \includegraphics[width=\figwid]{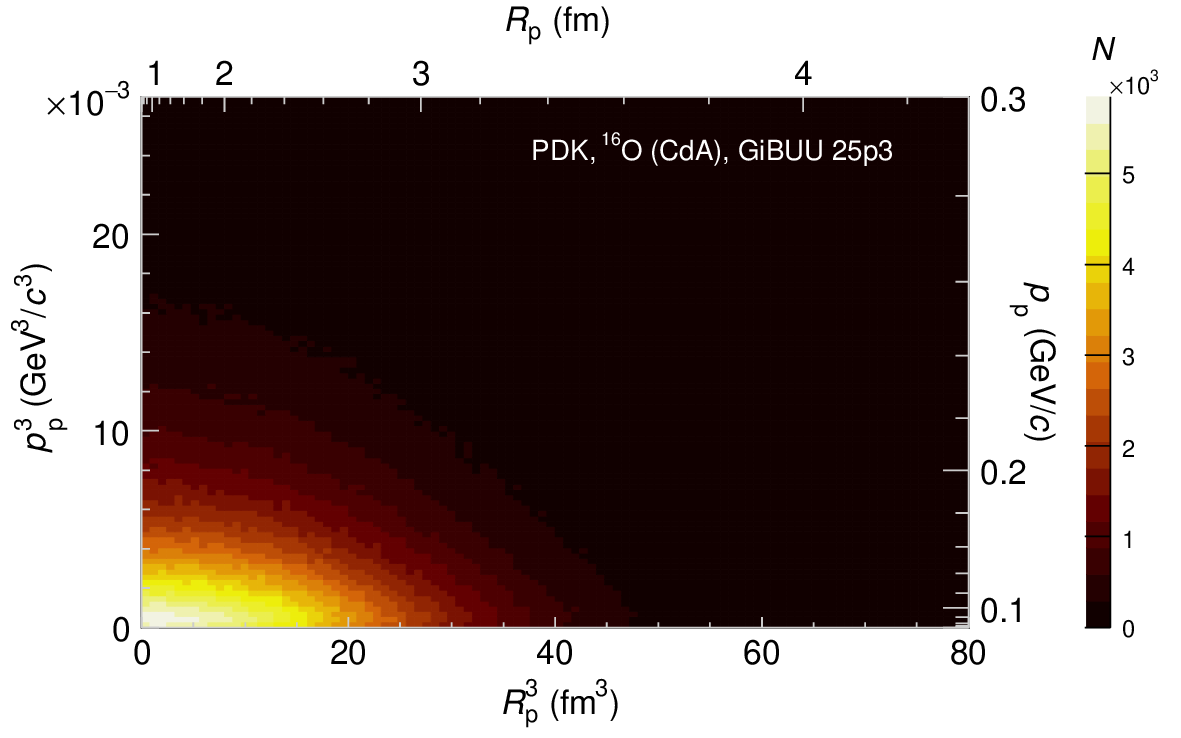}
        \caption{Phase space distribution.
        }\label{fig:r3p3b}
    \end{subfigure}
    \begin{subfigure}{\figwid}
        \centering
        \includegraphics[width=\figwid]{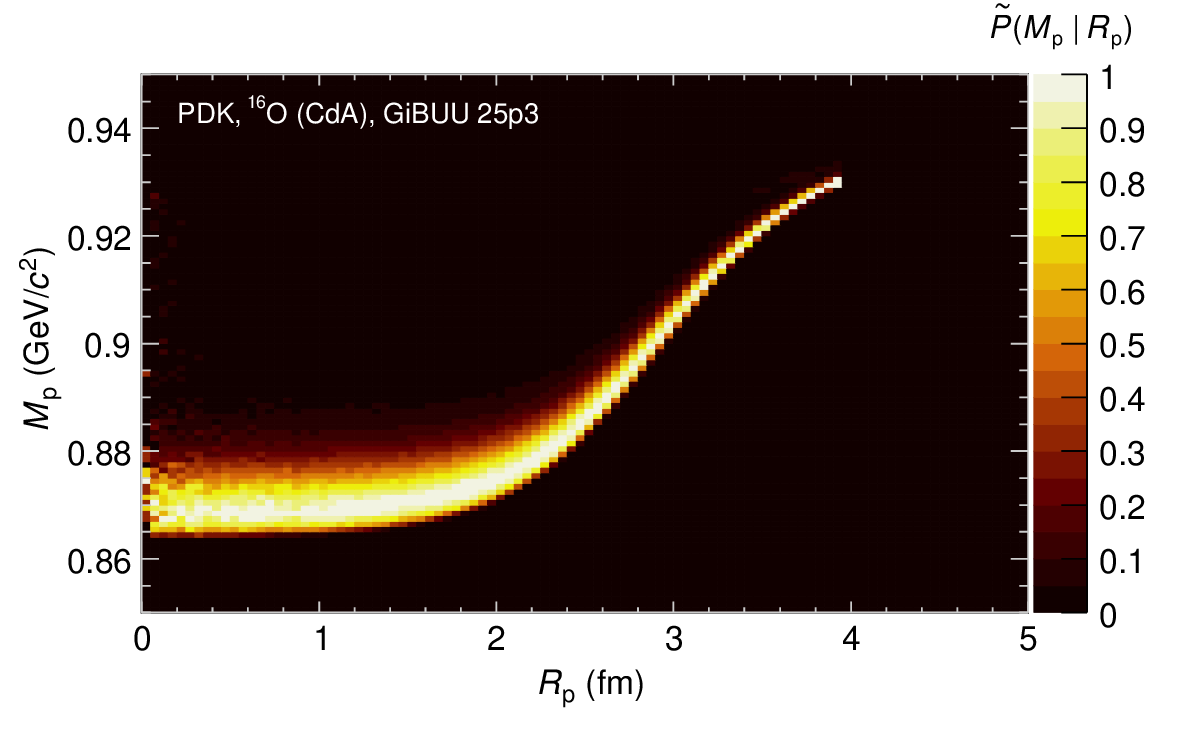}
        \caption{
            $\tilde{P}\left(M_\proton|R_\proton\right)$.
        }\label{fig:rMb}
    \end{subfigure}
    \begin{subfigure}{\figwid}
        \centering
        \includegraphics[width=\figwid]{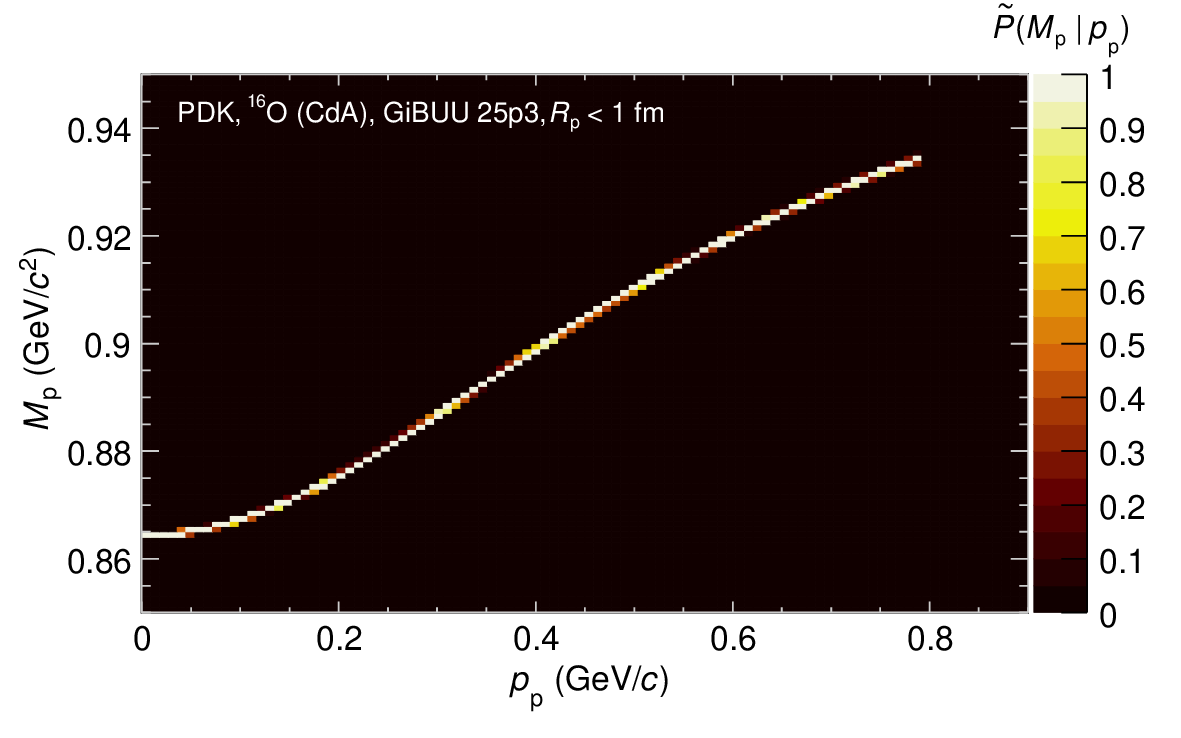}
        \caption{
            $\tilde{P}\left(M_\proton|p_\proton\right)$  for $R_\proton<1~\textrm{fm}$.
        }\label{fig:lfg_p3_mb}
    \end{subfigure}
    \caption{
        Similar to Fig.~\ref{fig:inirpma} but for \cda.
    }\label{fig:inirpmb}
\end{figure}

In addition to the LFG model in Fig.~\ref{fig:inirpma},
the $p_\proton$-$R_\proton$-$M_\proton$ correlations
with the \gibuu \cda model is shown in Fig.~\ref{fig:inirpmb}.

\newpage

\end{document}